\documentclass[journal]{IEEEtran}

\usepackage{amssymb,amsmath,graphicx,times,url,bm}
\usepackage{color}
\usepackage[printonlyused, nolist]{acronym}
\usepackage{algorithm,algorithmic}
\usepackage{pgfplots}
\usepackage{booktabs}
\usepackage{algorithm,algorithmic}
\usepackage{pifont}

\usepackage{tikz}
\usetikzlibrary{plotmarks}
\usetikzlibrary{calc}
\usetikzlibrary{arrows.meta}
\tikzset{%
	>={Latex[width=2mm,length=2mm]},
	base/.style = {rectangle, rounded corners, draw=black,
		minimum width=4cm, minimum height=1cm,
		text centered},
	activityStarts/.style = {base, fill=gray!30},
	startstop/.style = {base, fill=red!30},
	activityRuns/.style = {base, fill=green!30},
	process/.style = {base, minimum width=2.5cm},
}

\def\DrawCube#1#2#3{
	\begin{scope}[shift={#1}]
		\pgfmathsetmacro{\cubex}{0.3}
		\pgfmathsetmacro{\cubey}{3}
		\pgfmathsetmacro{\cubez}{0.3}
		\draw[{#2},fill={#3}] (0,0,0) -- ++(-\cubex,0,0) -- ++(0,-\cubey,0) -- ++(\cubex,0,0) -- cycle;
		\draw[{#2},fill={#3}] (0,0,0) -- ++(0,0,-\cubez) -- ++(0,-\cubey,0) -- ++(0,0,\cubez) -- cycle;
		\draw[{#2},fill={#3}] (0,0,0) -- ++(-\cubex,0,0) -- ++(0,0,-\cubez) -- ++(\cubex,0,0) -- cycle;
	\end{scope}
}

\def\DrawTileSquare#1#2#3#4{
	\begin{scope}[shift={#1}]
		\pgfmathsetmacro{\cubex}{3}
		\pgfmathsetmacro{\cubey}{0.2}
		\pgfmathsetmacro{\cubez}{3}
		\pgfmathsetmacro{\cubezBG}{1.95}
		\pgfmathsetmacro{\cubexBG}{1.75}
		\draw[{#2},fill={#3}] (0,0,0) -- ++(-\cubex,0,0) -- ++(0,-\cubey,0) -- ++(\cubex,0,0) -- cycle;
		\draw[{#2},fill={#3}] (0,0,0) -- ++(0,0,-\cubez) -- ++(0,-\cubey,0) -- ++(0,0,\cubez) -- cycle;
		\draw[{#2},fill={#3}] (0,0,0) -- ++(-\cubex,0,0) -- ++(0,0,-\cubez) -- ++(\cubex,0,0) -- cycle;
		\draw[{#2}] (0,0,0) -- ++(-\cubex,0,0) -- ++(0,0,-\cubezBG) -- ++(\cubex,0,0) -- cycle;
		\draw[{#2}] (0,0,0) -- ++(-\cubexBG,0,0) -- ++(0,0,-\cubezBG) -- ++(\cubexBG,0,0) -- cycle;
		\textcolor{black}{
			\node()at(-0.5,0.96){\small$\W^\text{SOI}_{{#4}}$};}
		\textcolor{black}{
			\node()at(-2,0.35){\small$\BGfilter_{{#4}}$};}
		\textcolor{black}{
			\node()at(-0.5,0.35){\small$-\mathbf{I}_{\MicIdxMax-\ChannelIdxMax}$};}
	\end{scope}
}

\pgfplotsset{compat=1.11}

\newlength\fwidth

\ifCLASSINFOpdf
\else
\fi
\newcommand{\MixingMat}{\mathbf{A}}
\newcommand{\SteeringVector}{\mathbf{h}}

\newcommand{\W}{\mathbf{W}}
\newcommand{\w}{\mathbf{w}}

\newcommand{\invW}{\mathbf{W}^{-1}}
\newcommand{\y}{\mathbf{y}}
\newcommand{\x}{\mathbf{x}}
\newcommand{\s}{\mathbf{s}}
\newcommand{\q}{\mathbf{q}}
\newcommand{\V}{\mathbf{V}}
\newcommand{\z}{\mathbf{z}}
\newcommand{\zTransf}{\tilde{\mathbf{z}}}
\newcommand{\BGvec}{\mathbf{b}}
\newcommand{\BGmat}{\mathbf{B}}
\newcommand{\BGvecTransf}{\tilde{\mathbf{b}}}
\newcommand{\BGmatTransf}{\tilde{\mathbf{B}}}
\newcommand{\BGfilter}{\BGmat^{\MicIdxMax,\ChannelIdxMax}}
\newcommand{\MicCorrMat}{\mathbf{C}}

\newcommand{\UpperBound}{U}

\newcommand{\EigMat}{\mathbf{T}_\FreqIdx}
\newcommand{\ScaleMat}{\mathbf{D}_\FreqIdx}

\newcommand{\BasisIdx}{\nu}
\newcommand{\BasisIdxMax}{N_\text{bases}}
\newcommand{\IterIdx}{l}
\newcommand{\IterIdxMax}{L}
\newcommand{\FreqIdx}{f}
\newcommand{\FreqIdxMax}{F}
\newcommand{\BlockIdx}{n}
\newcommand{\BlockIdxMax}{N}
\newcommand{\ChannelIdx}{k}
\newcommand{\ChannelIdxMax}{K}
\newcommand{\SrcIdx}{q}
\newcommand{\SrcIdxMax}{Q}
\newcommand{\MicIdx}{m}
\newcommand{\MicIdxMax}{M}
\newcommand{\sFreqVec}{\underline{\s}}
\newcommand{\yFreqVec}{\underline{\y}}
\newcommand{\xFreqVec}{\underline{\x}}
\newcommand{\zFreqVec}{\underline{\z}}

\newcommand{\SetW}{\mathcal{W}}

\newcommand{\Y}{\mathcal{Y}}
\newcommand{\X}{\mathcal{X}}
\newcommand{\Z}{\mathcal{Z}}

\newcommand{\SetN}{[N]}
\newcommand{\SetK}{[F]}

\newcommand{\SetIdxPrior}{\mathcal{I}}

\newcommand{\PriorCovMatInv}{\mathbf{P}_\FreqIdx^\ChannelIdx}
\newcommand{\PriorCovMatBG}{\mathbf{P}_\FreqIdx^{\text{BG}}}
\newcommand{\PriorCovMatNullInv}{\mathbf{P}_{\FreqIdx,\text{Null}}^\ChannelIdx}
\newcommand{\PriorCovMatOneInv}{\mathbf{P}_{\FreqIdx,\text{One}}^\ChannelIdx}

\newcommand{\transp}{^\text{T}}
\newcommand{\herm}{^\text{H}}
\newcommand{\inv}{^{-1}}
\newcommand{\Expect}[1]{\hat{\mathbb{E}}\left\lbrace #1 \right\rbrace}

\newcommand{\ILRMAVar}{\sigma^2}

\newcommand{\Walter}[1]{#1}
\newcommand{\WalterTwo}[1]{#1}
\newcommand{\WalterThree}[1]{#1}
\newcommand{\WalterFour}[1]{#1}

\begin{document}

\begin{acronym}
	\acro{STFT}{Short-Time Fourier Transform}
	\acro{PSD}{Power Spectral Density}
	\acro{PDF}{Probability Density Function}
	\acro{RIR}{Room Impulse Response}
	\acro{FIR}{Finite Impulse Response}
	\acro{FFT}{Fast Fourier Transform}
	\acro{DFT}{Discrete Fourier Transform}
	\acro{ICA}{Independent Component Analysis}
	\acro{IVA}{Independent Vector Analysis}
	\acro{TRINICON}{TRIple-N Independent component analysis for CONvolutive mixtures}
	\acro{FD-ICA}{Frequency Domain ICA}
	\acro{BSS}{Blind Source Separation}
	\acro{NMF}{Nonnegative Matrix Factorization}
	\acro{MM}{Majorize-Minimize}
	\acro{MAP}{Maximum A Posteriori}
	\acro{RTF}{Relative Transfer Function}
	\acro{auxIVA}{Auxiliary Function IVA}
	\acro{FD-ICA}{Frequency-Domain Independent Component Analysis}
	\acro{DOA}{Direction of Arrival}
	\acro{SNR}{Signal-to-Noise Ratio}
	\acro{SIR}{Signal-to-Interference Ratio}
	\acro{SDR}{Signal-to-Distortion Ratio}
	\acro{SAR}{Signal-to-Artefact Ratio}
	\acro{GC}{Geometric Constraint}
	\acro{DRR}{Direct-to-Reverberant energy Ratio}
	\acro{ILRMA}{Independent Low Rank Matrix Analysis}
	\acro{IVE}{Independent Vector Extraction}
	\acro{GC-IVA}{Geometric Constraint IVA}
	\acro{SOI}{Sources Of Interest}
	\acro{BG}{Background}
	\acro{MNMF}{Multichannel NMF}
	\acro{IP}{Iterative Projection}
\end{acronym}

\title{\Walter{A Unified Bayesian View on\\ Spatially Informed Source Separation and Extraction based on Independent Vector Analysis}}
%
%
%

\author{Andreas Brendel,~\IEEEmembership{Student Member,~IEEE,}
        Thomas Haubner, and~Walter~Kellermann,~\IEEEmembership{Fellow,~IEEE}
\thanks{\Walter{The authors} are with the chair of Multimedia Communications and Signal Processing, Friedrich-Alexander-Universit\"at Erlangen-N\"urnberg,
	Cauerstr. 7, D-91058 Erlangen, Germany,
	e-mail: \texttt{\{Andreas.Brendel, Thomas.Haubner, Walter.Kellermann\}@FAU.de}.}
\thanks{This work was supported by DFG under contract no $<$Ke890/10-1$>$ within the Research Unit FOR2457 "Acoustic Sensor Networks".}}

%
%

\markboth{}%
{Shell \MakeLowercase{\textit{et al.}}: Bare Demo of IEEEtran.cls for IEEE Journals}
%

\maketitle

\begin{abstract}
	Signal separation and extraction are important tasks for devices recording audio signals in real environments which\Walter{, aside from the desired sources,} often contain several interfering sources such as background noise or concurrent speakers. Blind Source Separation (BSS) provides a powerful \Walter{approach} to address such problems. However, \Walter{BSS algorithms} typically \Walter{treat} all sources \Walter{equally and do not resolve} uncertainty \Walter{regarding} the ordering of the \Walter{separated} signals at the output of the algorithm\Walter{, i.e., the outer permutation problem}. This paper addresses this problem by incorporating prior knowledge into the adaptation of the demixing filters\Walter{, e.g., the position of the sources,} in a Bayesian framework. We focus here on methods based on Independent Vector Analysis (IVA) \Walter{as it \WalterTwo{elegantly and successfully} deals with} the internal \Walter{permutation} problem. By including a background model, i.e., a model for sources we are not interested to separate, we enable the algorithm to extract the sources of interest in overdetermined and underdetermined scenarios at a low computational complexity. The proposed framework allows to \Walter{incorporate prior knowledge about the demixing filters in a generic way and unifies several known and newly proposed algorithms using a Bayesian view}. For all algorithmic variants, we provide efficient update rules based on the iterative projection principle. \Walter{The performance of a large variety of \WalterTwo{representative} algorithmic variants, \WalterTwo{including very recent algorithms}, is compared using measured \acp{RIR}.}
\end{abstract}


\IEEEpeerreviewmaketitle

\section{Introduction}
\IEEEPARstart{S}{ource} separation and signal extraction are essential \Walter{tasks} for \Walter{acoustic} signal processing on a variety of devices such as mobile phones, smart home assistants, hearing aids, conference systems etc. \Walter{For these tasks} many algorithms have been proposed in the recent years\Walter{, e.g.,} \cite{gannot_consolidated_2017,vincent_audio_2018} \Walter{which} can roughly \Walter{be} divided into two highly overlapping groups originating from different paradigms: beamforming methods \cite{van_veen_beamforming:_1988} and \ac{BSS} \Walter{\cite{vincent_audio_2018,hyvarinen_independent_2001,makino_blind_2007}}. 
In this paper, \Walter{we focus on the latter one}.

\Walter{As a first class of \ac{BSS} algorithms, we consider here algorithms which are based} on \ac{ICA} \cite{hyvarinen_independent_2001}, \cite{bell_information-maximization_1995}, \Walter{and} use the statistical independence of the source signals to derive algorithms capable of separating nongaussian sources. These methods are in general based on a linear instantaneous mixing and demixing model, which makes them not directly applicable for reverberant enclosures for which the recorded signals are filtered and superimposed versions of the source signals, \Walter{so that} a convolutive mixture model \Walter{should} be applied. As a solution, it has been proposed to apply the \ac{ICA} algorithm independently in \Walter{different frequency bins} \cite{smaragdis_blind_1998}. However, due to the well-known inner permutation problem, i.e., the uncertainty about the assignment of the demixed signals to the output channels in each frequency bin, the ordering of the channels has to be recovered by repair mechanisms \cite{sawada_robust_2004}. \Walter{For avoiding} the inner permutation problem, \ac{IVA} \cite{kim_blind_2007} has been introduced\Walter{, which enforces} statistical dependence between the frequency bins of the demixed signals. \Walter{For identifying the demixing system, stable}, fast and parameter-free update rules based on the \ac{MM} principle have been proposed in~\cite{ono_stable_2011}. 

Another class of algorithms for \WalterTwo{multichannel} source separation is based on \ac{MNMF} \cite{sawada_multichannel_2013}\Walter{, which is an extension of \ac{NMF}} \cite{lee_algorithms_2000}. The main idea here is to model the source signal spectrum by a superposition of nonnegative basis vectors. This approach is especially powerful if a distinct spectral structure can be exploited, e.g., for music signals \cite{fevotte_nonnegative_2009} or certain types of noise signals \cite{haubner_multichannel_2018}. 

An approach which \Walter{synthesizes} the ideas of \ac{IVA} and \ac{MNMF} has been \Walter{introduced as} \ac{ILRMA} \cite{kitamura_effective_2017,kitamura_determined_2016}. \ac{ILRMA} can either be understood as a special case of \ac{MNMF} using a \mbox{rank-1} spatial model or as \ac{IVA} with a time-varying Gaussian source model \cite{ono_user-guided_2012} whose variance is estimated via \ac{NMF}. The benefits of this approach are its faster convergence compared to \ac{MNMF} and the higher separation performance of sources with distinct \Walter{spectral} structure, e.g., music signals. However, if applied blindly, the permutation of the output channels remains arbitrary. Clustering based on the associated identified spatial models is \WalterTwo{difficult in a static and determined scenario, where the number of sources and sensors is equal. If the sources are moving or} the scenario is underdetermined, i.e., there are more sources than sensors, such \WalterTwo{a clustering-based method} is likely to fail.

For signal extraction\Walter{,} a \ac{BG} model has been proposed in \cite{koldovsky_gradient_2019} which \Walter{leads} to the \ac{IVE} algorithm. Here, one desired source is \Walter{separated} from a set of other sources forming the \ac{BG}, for which no effort is spent to separate them. The same model has been used in \cite{scheibler_2019} to derive an \ac{MM}-based optimization scheme for \ac{IVA} in overdetermined scenarios. In both cases it is argued that the coupling of the \ac{SOI} and the \ac{BG} is only weakly expressed in the cost function\Walter{, i.e., the cost function \WalterTwo{consists of} a part \WalterTwo{only} depending on the \ac{SOI} filters and another part \WalterTwo{only} depending on the \ac{BG} filters. As a remedy,} an orthogonality constraint is imposed on the demixing filters corresponding to \acp{SOI} and \ac{BG}, which yields the update rules for the \ac{BG} filters. For the selection of the \ac{SOI} filters, a directional constraint and \Walter{a} supervised adaptation based on a reference signal \cite{nesta_supervised_2017} has been suggested in \cite{kounovsky_recursive_2018} for \ac{IVE}. For \cite{scheibler_2019} no such selection strategy exists so far.

Many ways have been proposed to incorporate spatial prior knowledge about the sources into the adaptation of the demixing filters of \ac{BSS} algorithms to speed up convergence or to ensure the extraction of a desired source \Walter{\cite{parra_geometric_2002}}. A geometric constraint has \WalterTwo{also} been used in \ac{TRINICON}-based signal extraction \cite{yuanhang_zheng_bss_2009,reindl_minimum_2014,zheng_analysis_2014} \Walter{and for \ac{IVA}} in \cite{vincent_geometrically_2015}. An optimization algorithm for spatially regularized \ac{ILRMA} based on vector-wise coordinate descent has recently been proposed \WalterTwo{in} \cite{mitsui_vectorwise_2018}.

Besides geometric constraints, \cite{duong_under-determined_2010} proposed to use spatial models for the reverberant component of the observed sound signals together with free-field models to obtain a full-rank spatial covariance model. In \cite{koldovsky_semi-blind_2013}, previously obtained demixing filters are introduced as prior knowledge into \ac{BSS}.

In this paper, we propose a \Walter{novel} generic \WalterTwo{Bayesian} framework for informed source separation based on \ac{IVA}. \Walter{This framework allows to incorporate prior knowledge on the demixing matrices in a generic way and provides fast converging \ac{IP}-based update rules at a low computational complexity at the same time.} \WalterTwo{Various} \Walter{known and novel} algorithmic variants \WalterTwo{are identified as special cases of the generic framework}. Several \WalterTwo{strategies} for \WalterTwo{incorporating prior knowledge in the Bayesian sense are discussed and exemplified by} priors based on a free-field model, which allows to steer spatial ones and \WalterThree{nulls}. A \ac{BG} model \WalterThree{is introduced, which can also incorporate priors and allows for a significant reduction of computational cost.} For the \acp{SOI}, several source models \WalterTwo{are discussed} including \ac{NMF} and fast and stable update rules for all algorithmic variants based on the \ac{MM} principle \WalterTwo{are proposed}. \Walter{A new perspective is taken in the derivation of the update rules for the \ac{BG} filters based on \ac{IP}.} The proposed framework \Walter{allows} the solution of the outer permutation problem of \ac{BSS} as well as signal extraction and separation in determined and overdetermined \Walter{scenarios} and signal extraction in \WalterTwo{underdetermined} scenarios. \Walter{This paper is an extension of} \cite{brendel_2020}, \Walter{where we discussed a very specific realization of the generic \WalterTwo{Bayesian} framework presented here.}

In the following, scalar variables are typeset as lower-case letters, vectors as bold lower-case letters, matrices as bold upper-case letters and sets as calligraphic upper-case letters. $\mathbf{I}_d$ and $\mathbf{0}_d$ \Walter{denote} a quadratic identity or all-zero matrix, respectively, of dimensions $d\times d$\WalterTwo{,} and $\mathbf{0}_{d_1\times d_2}$ denotes \Walter{an} all-zero matrix of dimensions $d_1\times d_2$. $(\cdot)\herm$ and $(\cdot)\transp$ \Walter{denote} a Hermitian (complex conjugate transpose) and transposed matrix, respectively. Complex-conjugated quantities are marked by $(\cdot)^\ast$ \Walter{and the derivative of a function w.r.t. its argument is denoted by $(\cdot)'$}. The set $\{1,2,\dots,N\}$ is denoted by $[N]$. The notation of important variables is given in Tab.~\ref{tab:notations} for later reference. 
\begin{table}
	\centering
	\begin{tabular}{ll}
		\toprule
		$\mathbf{I}$, $\mathbf{0}$ &  Identity and all-zero matrix\\
		$\FreqIdx$, $\FreqIdxMax$  &  Frequency \WalterThree{bin} index and number of frequency bins\\
		$\ChannelIdx$, $\ChannelIdxMax$ & Channel index and number of channels \\
		$l$, $L$ &  Iteration index and number of iterations\\
		$\MicIdx$, $\MicIdxMax$ &  Microphone index and number of microphones\\
		$\BlockIdx$, $\BlockIdxMax$& Time block index and number of blocks \\
		$\BasisIdx$, $\BasisIdxMax$& Basis index and number of bases \\
		$\W$, $\w$ & Demixing matrix and demixing vector \\
		$\mathbf{P}$ &  Precision matrix of spatial prior\\
		$J$ & Cost function \\
		$\mathbf{A}$, $\mathbf{a}$ & Mixing matrix and mixing vector \\
		$t$, $v$& Basis element and activation of \ac{NMF}\\
		$\MicCorrMat$& Microphone covariance matrix of frequency bin $\FreqIdx$\\
		$\z$& \ac{BG} signal vector\\
		$\BGmat$, $\BGvec$ & \ac{BG} filter matrix and vector\\
		$Q$ & Number of sources\\
		$r$ & Estimated demixed signal variance \\
		$\UpperBound(\cdot\vert\cdot)$ & Upper bound\\
		$\V$ & Weighted microphone covariance matrix \\
		$\x$ & Microphone signal vector \\
		$\s$ & \ac{SOI} signal vector \\
		$\q$ & Source signal vector \\
		$\y$ & Demixed signal vector \\
		$\BGfilter$ & \ac{BG} filter submatrix \\
		$\yFreqVec$ & Broadband demixed signal vector \\
		$\zFreqVec$ & Broadband \ac{BG} signal vector \\
		$\Y, \X$ & Set of demixed signals and microphone signals \\
		$\SetW$ & Set of demixing matrices \\
		$\SetN, \SetK$ & Index set of time blocks and frequency bins \\
		$\SteeringVector_\FreqIdx(\vartheta)$ & Free-field steering vector for direction $\vartheta$\\
		\bottomrule
	\end{tabular}\vspace{1pt}
	\caption{Notations used}
	\label{tab:notations}
\end{table}

The remainder of the paper is structured as follows: Sec.~\ref{sec:model} defines the signal model, the probabilistic model for the \acp{SOI} and the \ac{BG} and introduces prior \acp{PDF}. The fundamental principle of \ac{MM} algorithms is \Walter{described} in Sec.~\ref{sec:MM_algorithm}. \Walter{In the same section}, an upper bound for the previously derived cost function is constructed \Walter{and} optimized\Walter{,} and update rules for the demixing filters based on the iterative projection principle are proposed. Experimental results are presented in Sec.~\ref{sec:experiments}. The paper is concluded in Sec.~\ref{sec:conslusion}.

\section{\Walter{Models}}
\label{sec:model}
The following section introduces the underlying \Walter{source} models for \acp{SOI} and \ac{BG} signals, the probabilistic model \Walter{for the demixing system including} prior \acp{PDF} which allow to incorporate prior knowledge about the demixing filters.
\subsection{Signal Model}
We consider an acoustic scene in an enclosure comprising $\MicIdxMax$ microphones and $\SrcIdxMax$ simultaneously active acoustic \Walter{point} sources observed by the microphones as a convolutive mixture. In this contribution, we are interested in separating $\ChannelIdxMax \leq \SrcIdxMax$ \acp{SOI} out of the observed mixture of $\SrcIdxMax$ sources. \Walter{The remaining sources, if there are any, are associated with the so-called \acf{BG} in the following.}

\Walter{With $\FreqIdx\in \SetK$ denoting the frequency \WalterTwo{bin} index and \mbox{$\BlockIdx\in \SetN$} the discrete time index, we assume a linear time-invariant} mixing model in the \ac{STFT} domain
\begin{equation}
\x_{\FreqIdx,\BlockIdx} = \MixingMat_\FreqIdx \q_{\FreqIdx,\BlockIdx},
\end{equation}
with the source signal vector
\begin{equation}
\q_{\FreqIdx,\BlockIdx} = \left[q_{1,\FreqIdx,\BlockIdx},\dots,q_{Q,\FreqIdx,\BlockIdx}\right]\transp \in \mathbb{C}^Q,
\end{equation}
the microphone signal vector
\begin{equation} \x_{\FreqIdx,\BlockIdx} = \left[x_{1,\FreqIdx,\BlockIdx},\dots,x_{\MicIdxMax,\FreqIdx,\BlockIdx}\right]\transp \in \mathbb{C}^{\MicIdxMax}
\end{equation}
and the mixing matrix containing the acoustic transfer functions \Walter{at frequency \WalterTwo{bin} $\FreqIdx$} from the source positions to the microphones
\begin{equation}
\MixingMat_\FreqIdx \in \mathbb{C}^{\MicIdxMax\times Q}.
\end{equation}
Note that the number of sources $Q$, the number of microphones $\MicIdxMax$ and the number of \acp{SOI} $\ChannelIdxMax$ can be different in general. 

\begin{figure}
	\scalebox{1}{
		\begin{tikzpicture}
		
		\coordinate (linksObenVorne) at (6,0); 
		\coordinate (linksUntenVorne) at (6,-3);
		\coordinate (linksObenHinten) at (6.75,1.05);
		
		\coordinate (mitteObenVorne) at (6,0);
		\coordinate (mitteObenHinten) at (6.75,1.05);
		
		\coordinate (rechtsObenVorne) at (12,0);
		\coordinate (rechtsObenHinten) at (12.75,1.05);
		
		\DrawCube{(mitteObenHinten)}{black}{lightgray};	
		\DrawCube{(mitteObenVorne)}{black}{lightgray};

		\DrawTileSquare{(10,-0.2)}{black}{lightgray}{1};
		\DrawTileSquare{(10,0)}{black}{lightgray}{1};
		\DrawTileSquare{(10,-2.8)}{black}{lightgray}{\FreqIdxMax};
		\node[rotate = 90]()at(9,-1){\small$\bullet\,\bullet\,\bullet$};
		
		\DrawCube{(rechtsObenHinten)}{black}{lightgray};
		\DrawCube{(12.45,0.6)}{black}{lightgray};

		\DrawCube{(12.3,0.45)}{black}{lightgray};	
		\DrawCube{(rechtsObenVorne)}{black}{lightgray};

		\draw[dotted, line width = 1pt] (5.7,0) -- (12,0);
		\draw[dotted, line width = 1pt] (5.7,-3) -- (12,-3);
		\draw[dotted, line width = 1pt] (5.7,-0.2) -- (12,-0.2);
		\draw[dotted, line width = 1pt] (5.7,-0.4) -- (12,-0.4);
		\draw[dotted, line width = 1pt] (5.7,-2.8) -- (12,-2.8);
		
		\node()at(5.3,-1.5){\small$\FreqIdx \downarrow$};
		
		\node()at(5.6,0.3){\small$\underline{\mathbf{x}}_{1,\BlockIdx}$};
		\node[rotate = 50]()at(6,0.8){\scalebox{0.4}{$\bullet\,\bullet\,\bullet$}};
		\node()at(6.4,1.4){\small$\underline{\mathbf{x}}_{\MicIdxMax,\BlockIdx}$};

		\node()at(11.3,0.4){$\bm{=}$};
		\node[rotate = 90]()at(11.3,-1){\small$\bullet\,\bullet\,\bullet$};
		\node()at(11.2,-2.5){$\bm{=}$};

		\node()at(7,0.4){$\bm{\times}$};
		\node[rotate = 90]()at(7,-1){\small$\bullet\,\bullet\,\bullet$};
		\node()at(6.9,-2.3){$\bm{\times}$};
		
		\node()at(12.1,-3.3){\small$\sFreqVec_{1,\BlockIdx}$};
		\node[rotate = 50]()at(12.31,-3.05){\scalebox{0.4}{$\bullet\,\bullet\,\bullet$}};
		\node()at(12.7,-2.8){\small$\sFreqVec_{\ChannelIdxMax,\BlockIdx}$};

		\node()at(12.9,-2.5){\small$\zFreqVec_{1,\BlockIdx}$};
		\node[rotate = 50]()at(13.15,-2.2){\scalebox{0.4}{$\bullet\,\bullet\,\bullet$}};
		\node()at(13.6,-1.9){\small$\zFreqVec_{\MicIdxMax-\ChannelIdxMax,\BlockIdx}$};
		\end{tikzpicture}}
	\caption{Illustration of the demixing process. The demixing matrix $\W_\FreqIdx$ is applied in each frequency bin separately to the broadband vectors of microphone signals $\underline{\mathbf{x}}_{\ChannelIdx,\BlockIdx}$, $\ChannelIdx\in[\MicIdxMax]$. The results are the extracted \acp{SOI} $\sFreqVec_{\ChannelIdx,\BlockIdx}$, $\ChannelIdx\in[\ChannelIdxMax]$\WalterTwo{,} and the \ac{BG} signals $\zFreqVec_{\ChannelIdx,\BlockIdx}$, $\ChannelIdx\in[\MicIdxMax-\ChannelIdxMax]$.}
	\label{fig:illustration_demixing}
\end{figure}
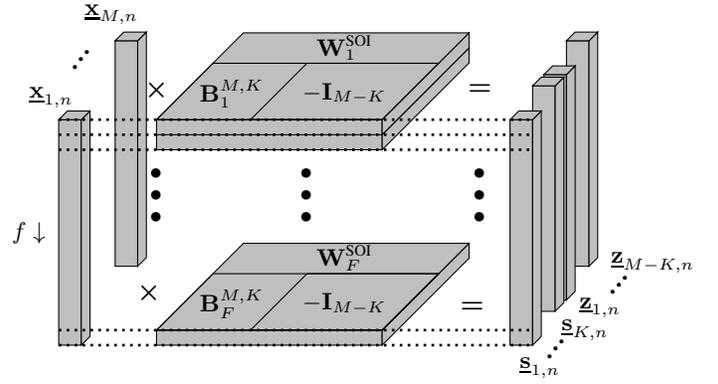

\WalterTwo{In the following, the demixing model is introduced as illustrated in Fig.~\ref{fig:illustration_demixing}.} The \acp{SOI} and the \ac{BG} signals are obtained by
\begin{equation}
\y_{\FreqIdx,\BlockIdx}  =\W_\FreqIdx \x_{\FreqIdx,\BlockIdx}
\label{eq:demixing_equation}
\end{equation}
\WalterTwo{where the} demixing matrix \Walter{applied in} frequency bin $\FreqIdx$
\begin{equation}
\W_\FreqIdx = \begin{bmatrix}
\W_\FreqIdx^{\text{SOI}}\\
\BGmat_\FreqIdx
\end{bmatrix}\in \mathbb{C}^{\MicIdxMax\times\MicIdxMax}
\end{equation}
contains two parts: One set of filters extracting the \acp{SOI} $\s_{\FreqIdx,\BlockIdx}$
\begin{equation}
\W_\FreqIdx^{\text{SOI}} = \begin{bmatrix}
\w_{\FreqIdx}^1,\dots,\w_\FreqIdx^\ChannelIdxMax
\end{bmatrix}\herm \in \mathbb{C}^{\ChannelIdxMax\times\MicIdxMax},
\end{equation}
and another \Walter{set of filters}
\begin{equation}
\BGmat_\FreqIdx = \begin{bmatrix}
\BGvec_\FreqIdx^1,\dots,\BGvec_\FreqIdx^{\MicIdxMax-\ChannelIdxMax}
\end{bmatrix}\herm =\begin{bmatrix}
\BGfilter_\FreqIdx&-\mathbf{I}_{\MicIdxMax-\ChannelIdxMax}
\end{bmatrix} \in \mathbb{C}^{\MicIdxMax-\ChannelIdxMax\times\MicIdxMax}
\label{eq:def_BG_matrix}
\end{equation}
estimating the \ac{BG} signals $\z_{\FreqIdx,\BlockIdx}$. Note that $\BGmat_\FreqIdx$ is \Walter{\WalterTwo{structured} according to the model proposed in \cite{koldovsky_gradient_2019}} \WalterTwo{with the identity matrix $\mathbf{I}_{\MicIdxMax-\ChannelIdxMax}$ and a submatrix $\BGfilter_\FreqIdx$ \WalterThree{capturing} the free parameters of $\BGmat_\FreqIdx$, which have to be identified together with the \ac{SOI} filters $\W_\FreqIdx^{\text{SOI}}$}. \WalterThree{For a given time frame $\BlockIdx$ and frequency bin $\FreqIdx$, the vector of output signals \WalterTwo{$\y_{\FreqIdx,\BlockIdx} = \begin{bmatrix}		\s_{\FreqIdx,\BlockIdx}\WalterThree{\transp},\z_{\FreqIdx,\BlockIdx}\WalterThree{\transp}\end{bmatrix}\transp$} contains the vector} of demixed \acp{SOI} denoted as
\begin{equation}
\s_{\FreqIdx,\BlockIdx} = \WalterTwo{\W_\FreqIdx^\text{SOI}\x_{\FreqIdx,\BlockIdx} =}  \left[s_{1,\FreqIdx,\BlockIdx},\dots,s_{\ChannelIdxMax,\FreqIdx,\BlockIdx}\right]\transp  \in \mathbb{C}^{\ChannelIdxMax}\WalterThree{,}
\end{equation}
and the vector of \ac{BG} signals denoted as
\begin{equation}
\z_{\FreqIdx,\BlockIdx} = \WalterTwo{\BGmat_\FreqIdx \x_{\FreqIdx,\BlockIdx} =} \left[z_{1,\FreqIdx,\BlockIdx},\dots,z_{\MicIdxMax-\ChannelIdxMax,\FreqIdx,\BlockIdx}\right]\transp \in \mathbb{C}^{\MicIdxMax-\ChannelIdxMax}.
\end{equation}
Note that only if $\ChannelIdxMax < \MicIdxMax$ holds, \ac{BG} signals can be \Walter{extracted by the assumed $\MicIdxMax\times\MicIdxMax$ demixing matrix $\W_\FreqIdx$}. 

For the determined case, i.e., $\ChannelIdxMax = \MicIdxMax$, no \ac{BG} signals are \Walter{estimated} and the demixing matrix separates only the \acp{SOI}
$\W_\FreqIdx = \W_\FreqIdx^{\text{SOI}}$. Furthermore, we define the broadband signal vector of the $\ChannelIdx$th \ac{SOI} and \ac{BG} signal \Walter{at time frame $\BlockIdx$}
\begin{equation*}
\sFreqVec_{\ChannelIdx,\BlockIdx} = \left[s_{\ChannelIdx,1,\BlockIdx},\dots,s_{\ChannelIdx,\FreqIdxMax,\BlockIdx}\right]\transp,\ \zFreqVec_{\ChannelIdx,\BlockIdx} = \left[z_{\ChannelIdx,1,\BlockIdx},\dots,z_{\ChannelIdx,\FreqIdxMax,\BlockIdx}\right]\transp  \in \mathbb{C}^{\FreqIdxMax}.
\end{equation*}
With the \Walter{definitions
\begin{equation}
\sFreqVec_{\BlockIdx} = \left[\sFreqVec_{1,\BlockIdx}\transp,\dots,\sFreqVec_{\ChannelIdxMax,\BlockIdx}\transp\right]\transp \in \mathbb{C}^{\ChannelIdxMax\FreqIdxMax}
\end{equation}
and}
\begin{equation}
\zFreqVec_{\BlockIdx} = \left[\zFreqVec_{1,\BlockIdx}\transp,\dots,\zFreqVec_{\MicIdxMax-\ChannelIdxMax,\BlockIdx}\transp\right]\transp \in \mathbb{C}^{(\MicIdxMax-\ChannelIdxMax)\FreqIdxMax}
\end{equation}
we can write the signal vector containing all output signals as
\begin{equation}
\yFreqVec_{\BlockIdx} = \left[\sFreqVec_{\BlockIdx}\transp,\zFreqVec_{\BlockIdx}\transp\right]\transp \in \mathbb{C}^{\MicIdxMax\FreqIdxMax}.
\end{equation}
Note that for the determined case\Walter{, i.e., $\MicIdxMax=\ChannelIdxMax$,}
\Walter{$\yFreqVec_{\BlockIdx} = \sFreqVec_{\BlockIdx}$} holds.
\subsection{Probabilistic Model \Walter{of the Demixing System}}
\Walter{For treating the identification of the demixing matrix as a Bayesian estimation problem, we} derive the posterior density of the demixing matrices \Walter{in the following}. Before \WalterTwo{starting} the derivation we define the set of all demixing matrices  \mbox{$\SetW = \left\lbrace\W_\FreqIdx \in \mathbb{C}^{\MicIdxMax\times\MicIdxMax}\vert \FreqIdx \in \SetK\right\rbrace$}, the set of all demixed signal vectors $\Y = \left\lbrace\yFreqVec_\BlockIdx \in \mathbb{C}^{\MicIdxMax \FreqIdxMax}\vert \BlockIdx \in \SetN\right\rbrace$ and the set of all microphone observations $\X = \left\lbrace\x_{\FreqIdx,\BlockIdx} \in \mathbb{C}^{\MicIdxMax}\vert \FreqIdx \in \SetK,\BlockIdx \in \SetN\right\rbrace$.

Using these definitions, the joint posterior of demixing matrices $\SetW$ and demixed signals $\Y$ can be written as
\begin{align}
p(\SetW,\Y\vert \X) &= p(\SetW,\Y) \frac{p(\X|\SetW,\Y)}{p(\X)}\notag\\
&\propto p(\SetW)p(\Y\vert\SetW) p(\X|\SetW,\Y).
\label{eq:joint_posterior}
\end{align}
We choose the following likelihood function for frequency bin $\FreqIdx$ and time step $\BlockIdx$, under the assumption that $\W_\FreqIdx$ is invertible
\begin{equation}
p\left(\x_{\FreqIdx,\BlockIdx} \big\vert \SetW,\y_{\FreqIdx,\BlockIdx}\right) = \delta \left(\x_{\FreqIdx,\BlockIdx}-\invW_\FreqIdx \y_{\FreqIdx,\BlockIdx}\right),
\label{eq:likelihood_binwise}
\end{equation}
where $\delta(\cdot)$ denotes the Dirac distribution. From \eqref{eq:likelihood_binwise} \Walter{a simplistic} likelihood for all frequency bins $\FreqIdx\in \SetK$ and time steps $\BlockIdx\in\SetN$ can be constructed by using an i.i.d. assumption
\begin{align}
p(\X|\SetW,\Y) &= \prod_{\BlockIdx=1}^{\BlockIdxMax}\prod_{\FreqIdx=1}^{\FreqIdxMax} \delta \left(\x_{\FreqIdx,\BlockIdx}-\invW_\FreqIdx \y_{\FreqIdx,\BlockIdx}\right).
\label{eq:likelihood}
\end{align}
\Walter{Moreover, a simplistic} probabilistic model for the sources can be \WalterTwo{formulated} under the assumption of independence between \WalterTwo{all time frames} as
\begin{equation}
p(\Y\vert\SetW) = \prod_{\BlockIdx=1}^{\BlockIdxMax}p\left(\yFreqVec_{\BlockIdx}\right) = \prod_{\BlockIdx=1}^{\BlockIdxMax}p\left(\zFreqVec_{\BlockIdx}\right)\prod_{\ChannelIdx=1}^{\ChannelIdxMax}p\left(\sFreqVec_{\ChannelIdx,\BlockIdx}\right),
\label{eq:source_model}
\end{equation}
\Walter{where in the rightmost term the realistic assumption of mutual statistical independence of the \acp{SOI} and the independence of the \acp{SOI} from the \ac{BG} sources is included.} Note that $p\left(\zFreqVec_{\BlockIdx}\right)$ and $p(\sFreqVec_{\ChannelIdx,\BlockIdx})$ are multivariate \acp{PDF} capturing all frequency bins. Now, the posterior of the demixing matrices is computed by marginalizing the demixed signals $\Y$ out of the joint posterior~\eqref{eq:joint_posterior}
\begin{align}
p(\SetW|\X) 
&\propto p(\SetW) \int p(\Y\vert\SetW) p(\X|\SetW,\Y) d\yFreqVec_1\dots d\yFreqVec_\BlockIdxMax.
\end{align}
Inserting the \Walter{models} \eqref{eq:likelihood} and \eqref{eq:source_model} yields
\begin{align}
p(\SetW|\X) &\propto 
p(\SetW) \prod_{\BlockIdx=1}^{\BlockIdxMax} \int p(\yFreqVec_{\BlockIdx}) \prod_{\FreqIdx=1}^{\FreqIdxMax} \delta \left(\x_{\FreqIdx,\BlockIdx}-\invW_\FreqIdx \y_{\FreqIdx,\BlockIdx}\right) d\yFreqVec_\BlockIdx. \notag
\end{align}
Applying the rules for a linear transform of complex random variables \Walter{\cite{moreau_blind_2013}} \Walter{\WalterTwo{to} the transform $\yFreqVec_{\FreqIdx,\BlockIdx} = \W_\FreqIdx\xFreqVec_{\FreqIdx,\BlockIdx}$ and \WalterTwo{using} the sifting property of the Dirac distribution} yields finally
\begin{align}
p(\SetW|\X) & \propto
p(\SetW) \prod_{\FreqIdx=1}^{\FreqIdxMax} \vert \det \W_\FreqIdx \vert^{2\BlockIdxMax}\prod_{\BlockIdx=1}^{\BlockIdxMax}p\left(\zFreqVec_{\BlockIdx}\right)\prod_{\ChannelIdx=1}^{\ChannelIdxMax} p(\sFreqVec_{\ChannelIdx,\BlockIdx}).
\label{eq:posterior_W}
\end{align}
Optimizing the posterior for the demixing matrices \Walter{considering the logarithm of \eqref{eq:posterior_W}} yields the following \ac{MAP} problem
\begin{align}
\SetW &= \underset{\SetW}{\arg\max}\,\frac{\log p(\SetW)}{\BlockIdxMax} + 2\sum_{\FreqIdx=1}^{\FreqIdxMax} \log \vert \det \W_\FreqIdx \vert \dots\notag\\
& \quad \dots - \sum_{\ChannelIdx=1}^{\ChannelIdxMax}\Expect{G\left(\sFreqVec_{\ChannelIdx,\BlockIdx}\right)} + \Expect{\log p\left(\zFreqVec_{\BlockIdx}\right)}.
\label{eq:MAP_problem}
\end{align}
Here, we introduced the \Walter{score function} \mbox{$G(\sFreqVec_{\ChannelIdx,\BlockIdx}) = -\log p(\sFreqVec_{\ChannelIdx,\BlockIdx})$} and the averaging operator
$\Expect{\cdot} = \frac{1}{\BlockIdxMax} \sum_{\BlockIdx=1}^{\BlockIdxMax}(\cdot)$ for a concise notation.

\subsection{Models for \acp{SOI}}
\label{sec:foreground_models}
In the following, we want to introduce \Walter{various} widely-used models $p(\sFreqVec_{\ChannelIdx,\BlockIdx})$ for the \acp{SOI}.
\subsubsection{Super-Gaussian \ac{PDF}}
A popular and flexible source model for \ac{IVA}, containing many others as a special case, is the generalized Gaussian distribution \cite{ono_auxiliary-function-based_2012}
\begin{equation}
p\left(\sFreqVec_{\ChannelIdx,\BlockIdx}\right) \propto\exp\left(-\Vert\sFreqVec_{\ChannelIdx,\BlockIdx}\Vert_2^\beta\right),
\label{eq:GGD_PDF}
\end{equation}
where $\beta\in\mathbb{R}_+$ the shape parameter and $\Vert\cdot\Vert_2$ the Euclidean norm. The corresponding score function is given as (discarding constant terms)
\begin{equation}
	G(\sFreqVec_{\ChannelIdx,\BlockIdx}) = \Vert \sFreqVec_{\ChannelIdx,\BlockIdx} \Vert_2^\beta.
\end{equation}
\subsubsection{Time-varying Gaussian \ac{PDF}}
A Gaussian \ac{PDF} with time-varying broadband signal variance $\sigma^2_{\ChannelIdx,\BlockIdx}$~\Walter{\cite{ono_auxiliary-function-based_2012}}
\begin{equation}
	p\left(\sFreqVec_{\ChannelIdx,\BlockIdx}\right) \propto \exp\left(-\frac{\Vert\sFreqVec_{\ChannelIdx,\BlockIdx}\Vert_2^2}{\sigma^2_{\ChannelIdx,\BlockIdx}}\right),
	\label{eq:SOI_model_timevaryingGaussian}
\end{equation}
\WalterTwo{is another popular choice,} where the corresponding score function is given as (discarding constant terms)
\begin{equation}
	G(\sFreqVec_{\ChannelIdx,\BlockIdx}) = \frac{\Vert\sFreqVec_{\ChannelIdx,\BlockIdx}\Vert_2^2}{\sigma^2_{\ChannelIdx,\BlockIdx}}.
\end{equation}
\subsubsection{Nonnegative Matrix Factorization}
\label{sec:NMF_sourceModel}
If the source signal spectrum is structured, e.g., for music signals, or if prior knowledge about the source spectrum is available, an \ac{NMF}-based source model is \Walter{promising}. \Walter{Hereby,} independence over all frequency bins is assumed \cite{kitamura_effective_2017}
\begin{equation}
p\left(\sFreqVec_{\ChannelIdx,\BlockIdx}\right) = \prod_{\FreqIdx =1}^{\FreqIdxMax} \mathcal{N}^C\left(s_{\ChannelIdx,\FreqIdx,\BlockIdx}\vert0,\ILRMAVar_{\ChannelIdx,\FreqIdx,\BlockIdx}\right)
\end{equation}
\WalterTwo{where the} circularly-symmetric complex Gaussian distribution 
\begin{equation}
\mathcal{N}^C\left(s_{\ChannelIdx,\FreqIdx,\BlockIdx}\vert0,\ILRMAVar_{\ChannelIdx,\FreqIdx,\BlockIdx}\right) = \frac{1}{\pi \ILRMAVar_{\ChannelIdx,\FreqIdx,\BlockIdx}}\exp\left(-\frac{\vert s_{\ChannelIdx,\FreqIdx,\BlockIdx}\vert^2}{\ILRMAVar_{\ChannelIdx,\FreqIdx,\BlockIdx}}\right)
\end{equation}
for each time-frequency bin has been \WalterTwo{chosen} \cite{kitamura_determined_2016}. 
The \Walter{frequency} bin-wise signal variance $\ILRMAVar_{\ChannelIdx,\FreqIdx,\BlockIdx} = \mathbb{E}\{\vert s_{\ChannelIdx,\FreqIdx,\BlockIdx}\vert^2\}$ is modeled as
\begin{equation}
\hat{\sigma}^2_{\ChannelIdx,\FreqIdx,\BlockIdx} = \left(\sum_{\BasisIdx=1}^{\BasisIdxMax}t_{\ChannelIdx,\FreqIdx,\Walter{\BasisIdx}}v_{\ChannelIdx,\Walter{\BasisIdx},\BlockIdx}\right)^\beta,
\label{eq:NMF_r}
\end{equation}
where $\beta\in\mathbb{R}_+$ is \Walter{a} user-defined parameter. Hereby, \mbox{$\BasisIdx\in[\BasisIdxMax]$} indexes the basis vectors, $t_{\ChannelIdx,\FreqIdx,\BasisIdx}$ denotes the element of the $\BasisIdx$th basis vector corresponding to frequency bin $\FreqIdx$ and source $\ChannelIdx$ and the associated activation at time instant $\BlockIdx$ is denoted by $v_{\ChannelIdx,\BasisIdx,\BlockIdx}$. The resulting score function reads (discarding constant terms)
\begin{equation}
G\left(\sFreqVec_{\BlockIdx}\right) = \sum_{\FreqIdx=1}^{\FreqIdxMax}\sum_{\ChannelIdx=1}^{\ChannelIdxMax}\left(\log\ILRMAVar_{\ChannelIdx,\FreqIdx,\BlockIdx} + \frac{\vert s_{\ChannelIdx,\FreqIdx,\BlockIdx}\vert^2}{\ILRMAVar_{\ChannelIdx,\FreqIdx,\BlockIdx}}\right).
\end{equation}
An in-depth discussion of different source models \Walter{for \ac{ILRMA}}, \WalterTwo{where \ac{NMF} source models are commonly used,} can be found in \cite{kitamura_generalized_2018}.
\subsection{Background Model}
\label{sec:background_model}
We  model the \ac{BG} signals\Walter{, collected in set \mbox{$\Z = \left\lbrace\z_{\FreqIdx,\BlockIdx} \in \mathbb{C}^{\MicIdxMax}\vert \FreqIdx \in \SetK,\BlockIdx \in \SetN\right\rbrace$},} to be independent over all frequency bins and time steps for simplicity 
\begin{equation}
p(\Z) =  \prod_{\BlockIdx=1}^{\BlockIdxMax}p\left(\zFreqVec_{\BlockIdx}\right) = \prod_{\BlockIdx=1}^{\BlockIdxMax}\prod_{\FreqIdx=1}^{\FreqIdxMax}p\left(\z_{\FreqIdx,\BlockIdx}\right).
\label{eq:BG_IID_assumption}
\end{equation}
Furthermore, we model the \ac{BG} signals at each time-frequency bin to be multivariate complex Gaussian distributed
\begin{equation}
	p\left(\z_{\FreqIdx,\BlockIdx}\right) = \frac{1}{\pi^{\MicIdxMax-\ChannelIdxMax}\vert\det\mathbf{R}_\FreqIdx\vert}\exp\left(-\z_{\FreqIdx,\BlockIdx}\herm\mathbf{R}_\FreqIdx\inv\z_{\FreqIdx,\BlockIdx}\right),
	\label{eq:BG_modelPDF}
\end{equation}
where $\mathbf{R}_\FreqIdx$ denotes its covariance matrix. Note that we do not aim \WalterThree{at separating} the \ac{BG} signals and neither aim \WalterThree{at estimating} their covariance matrix. \WalterThree{Note that \eqref{eq:BG_modelPDF} puts no restrictions on the \ac{BG} model except for Gaussianity, \WalterFour{so} that, e.g., spatially white noise as well as spatially correlated sound fields, notably diffuse sound fields, are captured.}

To simplify the derivation of the update algorithms for the \ac{BG} filters, we \WalterThree{use} an eigenvalue decomposition of the \ac{BG} signal covariance matrix
\begin{equation}
	\EigMat\herm\mathbf{R}_\FreqIdx\inv\EigMat = \bm{\Lambda}_\FreqIdx.
	\label{eq:EVD_BG}
\end{equation}
Hereby, $\EigMat \in \mathbb{C}^{(\MicIdxMax-\ChannelIdxMax)\times(\MicIdxMax-\ChannelIdxMax)}$ denotes an orthonormal matrix (i.e., $\EigMat\EigMat\herm = \mathbf{I}_{\MicIdxMax-\ChannelIdxMax}$) containing the eigenvectors of $\mathbf{R}_\FreqIdx$ and $\bm{\Lambda}_\FreqIdx$ denotes a diagonal matrix containing its eigenvalues. Note that such a decomposition always \Walter{exists} for covariance matrices. As all eigenvalues are real-valued and positive, $\bm{\Lambda}_\FreqIdx$ can be decomposed as
\begin{equation}
	\bm{\Lambda}_\FreqIdx = \ScaleMat\ScaleMat,
	\label{eq:scaling_BG}
\end{equation}
where $\ScaleMat\in \mathbb{R}^{(\MicIdxMax-\ChannelIdxMax)\times(\MicIdxMax-\ChannelIdxMax)}$ denotes the matrix square root of $\bm{\Lambda}_\FreqIdx$. Note that the entries of $\ScaleMat$ are again all real-valued and positive, hence, $\ScaleMat$ is invertible.

Using the relations \eqref{eq:EVD_BG} and \eqref{eq:scaling_BG}, the covariance matrix $\mathbf{R}_\FreqIdx$ can be transformed into an identity matrix
\begin{equation}
\ScaleMat\inv\EigMat\herm\mathbf{R}_\FreqIdx\inv\EigMat\ScaleMat\inv = \mathbf{I}_{\MicIdxMax-\ChannelIdxMax}.
\label{eq:covMat2Identity}
\end{equation}
By using \eqref{eq:covMat2Identity}, we obtain
\begin{align}
p\left(\z_{\FreqIdx,\BlockIdx}\right)&=\frac{1}{\pi^{\MicIdxMax-\ChannelIdxMax}\vert\det\mathbf{R}_\FreqIdx\vert}\exp\left(-\zTransf_{\FreqIdx,\BlockIdx}\herm\zTransf_{\FreqIdx,\BlockIdx}\right),
\end{align}
with
\begin{equation}
	\zTransf_{\FreqIdx,\BlockIdx} = \ScaleMat\EigMat\herm\z_{\FreqIdx,\BlockIdx} = \ScaleMat\EigMat\herm\BGmat_\FreqIdx\x_{\FreqIdx,\BlockIdx} = \BGmatTransf_\FreqIdx\x_{\FreqIdx,\BlockIdx}.
	\label{eq:BG_transformation}
\end{equation}
Here, we defined $\BGmatTransf_\FreqIdx = \ScaleMat\EigMat\herm\BGmat_\FreqIdx$. Taking the i.i.d. assumption \eqref{eq:BG_IID_assumption} w.r.t. time and frequency of the \ac{BG} signals into account, the \ac{PDF} of all \ac{BG} signals $\Z$ is obtained as
\begin{align}
p(\Z) &\propto\exp\left(-\sum_{\FreqIdx=1}^{\FreqIdxMax}\sum_{\BlockIdx=1}^{\BlockIdxMax}\zTransf_{\FreqIdx,\BlockIdx}\herm\zTransf_{\FreqIdx,\BlockIdx}\right)\\
&=\exp\left(-\sum_{\FreqIdx=1}^{\FreqIdxMax}\sum_{\BlockIdx=1}^{\BlockIdxMax}\sum_{\ChannelIdx=1}^{\MicIdxMax-\ChannelIdxMax} (\BGvecTransf_\FreqIdx^\ChannelIdx)\herm\x_{\FreqIdx,\BlockIdx}\x_{\FreqIdx,\BlockIdx}\herm\BGvecTransf_\FreqIdx^\ChannelIdx\right)\\
&=\exp\left(-\BlockIdxMax\sum_{\FreqIdx=1}^{\FreqIdxMax}\sum_{\ChannelIdx=1}^{\MicIdxMax-\ChannelIdxMax} (\BGvecTransf_\FreqIdx^\ChannelIdx)\herm\MicCorrMat_\FreqIdx\BGvecTransf_\FreqIdx^\ChannelIdx\right).
\end{align}
Hereby, $\BGvecTransf_\FreqIdx^\ChannelIdx$ denote the modified \ac{BG} filter vectors, defined analogously to \eqref{eq:def_BG_matrix} and $\MicCorrMat_\FreqIdx = \Expect{\x_{\FreqIdx,\BlockIdx}\x_{\FreqIdx,\BlockIdx}\herm}$ the microphone signal covariance matrix. Hence, we obtain the following term contributing to the cost function (neglecting constant terms)
\begin{equation}
\log p(\Z) = -\BlockIdxMax\sum_{\FreqIdx=1}^{\FreqIdxMax}\sum_{\ChannelIdx=1}^{\MicIdxMax-\ChannelIdxMax} (\BGvecTransf_\FreqIdx^\ChannelIdx)\herm\MicCorrMat_\FreqIdx\BGvecTransf_\FreqIdx^\ChannelIdx = -\BlockIdxMax J_\text{BG}(\SetW).
\label{eq:BG_costFctTerm}
\end{equation}
\subsection{Priors}
\label{sec:priors}
The prior of the demixing matrices is chosen to be the product of marginal \acp{PDF} for each \ac{SOI} filter $\w_\FreqIdx^\ChannelIdx$, the \ac{BG} filter matrix $\BGmat_\FreqIdx$ and frequency bin $\FreqIdx$ 
\begin{equation}
p(\SetW) = \prod_{\FreqIdx=1}^{\FreqIdxMax}p(\W_\FreqIdx)= \prod_{\FreqIdx=1}^{\FreqIdxMax}p\left(\BGmat_\FreqIdx\right)\prod_{\ChannelIdx\in\SetIdxPrior}p\left(\w_\FreqIdx^\ChannelIdx\right).
\label{eq:prior_PDF}
\end{equation}
In the following, we will discuss separately the priors for the \ac{SOI} and the \ac{BG} filters and will give the overall term contributing to the cost function. 
\subsubsection{\acp{SOI}}
In many cases no prior knowledge is available for some of the channels or the optimization of the corresponding demixing filters should not be constrained. Hence, we only incorporate prior knowledge for a subset $\SetIdxPrior\subseteq[\ChannelIdxMax]$ of the demixing filters of the \acp{SOI} and choose uninformative priors for $\ChannelIdx\notin\SetIdxPrior$. In the following, we will present two different priors for the \ac{SOI} filters based on Gaussian \acp{PDF}.

The first option for a prior for the $\ChannelIdx$-th channel is chosen to be a zero-mean complex multivariate Gaussian \ac{PDF} with precision matrix $\PriorCovMatInv$ and weighting factor $\tilde{\gamma}_{\ChannelIdx,\FreqIdx}$
\begin{equation}
p\left(\w_\FreqIdx^\ChannelIdx\right) = \frac{\sqrt{(\tilde{\gamma}_{\ChannelIdx,\FreqIdx})^\MicIdxMax\det \PriorCovMatInv}}{\sqrt{\pi^\MicIdxMax}}\exp\left(-\tilde{\gamma}_{\ChannelIdx,\FreqIdx} (\w_\FreqIdx^\ChannelIdx)\herm\PriorCovMatInv\w_\FreqIdx^\ChannelIdx\right).
\label{eq:directional_prior}
\end{equation}
The weighting factor $\tilde{\gamma}_{\ChannelIdx,\FreqIdx}$ controls here and similarly for the other priors the impact of the prior on the overall model, i.e., it is a user-defined parameter. In the following, we want to discuss different choices for $\PriorCovMatInv$ yielding different priors for the demixing filters. 
To construct these priors, we use a free-field model and define the steering vector as
\begin{equation}
[\SteeringVector_\FreqIdx(\vartheta_i)]_\MicIdx = \left[\exp\left(j\frac{2\pi \mu_\FreqIdx}{c_s}\Vert\mathbf{r}_\MicIdx-\mathbf{r}_1\Vert_2\cos \vartheta_i\right)\right]_\MicIdx,
\label{eq:def_steering_vector}
\end{equation}
where $\mathbf{r}_\MicIdx$ denotes the position of the $\MicIdx$th microphone, $\mu_\FreqIdx$ the frequency in $\mathrm{Hz}$ corresponding to frequency bin $\FreqIdx$, $\vartheta_i$ the direction of the source and $c_s$ the speed of sound. Using this definition, we define the precision matrix yielding a spatial null
\begin{equation}
	\PriorCovMatNullInv = \lambda^\text{Null}_\text{Tik}\mathbf{I}_\MicIdxMax + \sum_{i:\vartheta_i\in\Theta_\ChannelIdx}\lambda^\text{Null}_i\mathbf{h}_\FreqIdx(\vartheta_i)\mathbf{h}_\FreqIdx(\vartheta_i)\herm,
	\label{eq:precision_matrix_null}
\end{equation}
where $\Theta_\ChannelIdx$ denotes the set of constrained \acp{DOA} and $\lambda^\text{Null}_i$ is a weight defining the influence of the constraint in direction $\vartheta_i$\WalterTwo{, while} \Walter{$\lambda^\text{Null}_\text{Tik}$ controls} the penalty on the filters energy. The intuition behind this choice can be understood if the argument of \eqref{eq:directional_prior} is rearranged
\begin{align}
	&(\w_\FreqIdx^\ChannelIdx)\herm\PriorCovMatNullInv\w_\FreqIdx^\ChannelIdx = \cdots\\ 
	&\qquad \cdots=\lambda^\text{Null}_\text{Tik} \Vert\w_\FreqIdx^\ChannelIdx\Vert_2^2 + \sum_{i:\vartheta_i\in\Theta_\ChannelIdx}\lambda^\text{Null}_i\Vert\mathbf{h}_\FreqIdx(\vartheta_i)\herm\w_\FreqIdx^\ChannelIdx \Vert_2^2\nonumber.
\end{align}
The first term represents the filters power and can be seen as a Tikhonov regularizer. The second term gives the length of the projection of the filters \Walter{$\w_\FreqIdx^\ChannelIdx$} onto the steering vectors \WalterTwo{$\mathbf{h}_\FreqIdx(\vartheta_i)$}. Hence, this prior favors solutions with small filter energy and \Walter{good angular alignment} to the steering vectors \Walter{$\mathbf{h}_\FreqIdx(\vartheta_i)$}. Similarly, the precision matrix yielding a spatial one is given as
\begin{equation}
\PriorCovMatOneInv = \lambda^\text{One}_\text{Tik}\mathbf{I}_\MicIdxMax - \sum_{i:\vartheta_i\in\Theta_\ChannelIdx}\lambda^\text{One}_i\mathbf{h}_\FreqIdx(\vartheta_i)\mathbf{h}_\FreqIdx(\vartheta_i)\herm,
\label{eq:precision_matrix_one}
\end{equation}
where $\lambda^\text{One}_i$ \Walter{and $\lambda^\text{One}_\text{Tik}$} are weighting parameters.

As an alternative to \eqref{eq:directional_prior}, we present another prior for the channels $\ChannelIdx\in\mathcal{I}^\text{Euc}$ based on the Euclidean distance between the current filter estimate and the target filter vector
\begin{align}
	&p\left(\w_\FreqIdx^\ChannelIdx\right) =  \frac{\sqrt{(\tilde{\gamma}_{\ChannelIdx,\FreqIdx}^{\text{Euc}})^\MicIdxMax}}{\sqrt{\pi^\MicIdxMax}}\dots \label{eq:directional_prior_Eucl}\\
	&\qquad\dots \exp\left(-\tilde{\gamma}_{\ChannelIdx,\FreqIdx}^{\text{Euc}}(\w_\FreqIdx^\ChannelIdx - \SteeringVector_\FreqIdx(\vartheta_\ChannelIdx))\herm(\w_\FreqIdx^\ChannelIdx - \SteeringVector_\FreqIdx(\vartheta_\ChannelIdx))\right).\notag
\end{align}
Hereby, we used the the steering vector $\SteeringVector_\FreqIdx(\vartheta_\ChannelIdx)$ defined in~\eqref{eq:def_steering_vector}.

In this contribution, we discuss practical realizations of the priors on the demixing vectors in the form of spatial priors which will also be the main focus in this paper. However, it should be noted that the proposed framework can be used for any prior which can be represented in the form of \eqref{eq:directional_prior} or \eqref{eq:directional_prior_Eucl}. \Walter{Note that \eqref{eq:precision_matrix_null} and \eqref{eq:directional_prior_Eucl} have been first introduced in \cite{mitsui_vectorwise_2018} and \cite{brendel_2020}, respectively.}
\subsubsection{Background}
Analogously to the priors for the \acp{SOI} \eqref{eq:directional_prior}, we choose the prior for the transformed \ac{BG} filters to be
\begin{align}
&p\left(\BGmatTransf_\FreqIdx\right) = \left(\frac{\sqrt{(\tilde{\gamma}_\FreqIdx^\text{BG})^\MicIdxMax\det \PriorCovMatBG}}{\sqrt{\pi^\MicIdxMax}}\right)^{\MicIdxMax-\ChannelIdxMax} \dots \notag\\
&\qquad\dots\exp\left(- \tilde{\gamma}_\FreqIdx^\text{BG}\sum_{\ChannelIdx=1}^{\MicIdxMax-\ChannelIdxMax}(\BGvecTransf_\FreqIdx^{\ChannelIdx})\herm\PriorCovMatBG\BGvecTransf_\FreqIdx^{\ChannelIdx}\right),
\label{eq:directional_prior_BG}
\end{align}
where we assumed independence between all channels and impose the same constraint by choosing $\PriorCovMatBG$ according to \eqref{eq:precision_matrix_null} for all \ac{BG} channels. \WalterTwo{Note that the independence assumption \WalterThree{applies here to} the filters, not \WalterThree{to} the \ac{BG} signals. This can be justified by considering filters associated with independent source positions to be independent as well.} The constrained directions for the \ac{BG} are collected in the set $\Theta_\text{BG}$. \WalterTwo{\WalterThree{Thereby}, one or multiple spatial nulls can be controlled, e.g., to avoid the occurrence of the \acp{SOI} in the \ac{BG}.}

\subsubsection{Overall Prior}
\begin{figure}
	\centering
	\scalebox{0.9}{
	\begin{tikzpicture}[node distance=1.5cm,
	every node/.style={fill=white, font=\sffamily}, align=center]
	\node (micSignals)        [activityStarts]{Microphone signals $\X$};
	\node (sourceModel)       [process, below of=micSignals]{Source Model};
	\node (BSS)       [process, below of=sourceModel, yshift=-1cm]{BSS};
	\node (BG)       [process, left of=BSS, xshift=-1.5cm]{Signal Extraction};
	\node (SigExtr)       [process, right of=BSS, xshift=1.5cm]{BG};
	\node (Output)       [activityStarts, below of=BSS, yshift=-1cm, xshift=-1cm]{Demixed Signals $\Y$};
	\node (directions) [process, below of=SigExtr, yshift=-0.5cm]{Prior Knowledge};
	
	\draw[->]        (micSignals) -- (sourceModel);
	\draw[->]       (sourceModel) -- node {$\ChannelIdxMax=\MicIdxMax$}(BSS);
	\draw[->]       (sourceModel) -- node [xshift = -0.75cm]{$Q>\MicIdxMax$,$\ChannelIdxMax\leq\MicIdxMax$}(BG);
	\draw[->]       (sourceModel) -- node {$\ChannelIdxMax<\MicIdxMax$}(SigExtr);
	\draw[->]       (BG) -- (Output);
	\draw[->]       (BSS) -- (Output);
	\draw[->]       (SigExtr) -- (Output);
	\draw[->]       (directions) -- node {$\Theta_\ChannelIdx$}(BG);
	\draw[->]       (directions) -- node {$\Theta_\ChannelIdx$}(BSS);
	\draw[->]       (directions) -- node {$\Theta_\text{BG}$}(SigExtr);
	\end{tikzpicture}}
	\caption{Relation of proposed algorithmic variants. Depending on $\SrcIdxMax$, $\ChannelIdxMax$ and $\MicIdxMax$, different algorithmic variants can be chosen: determined source separation, signal extraction or overdetermined \ac{BSS} using a \ac{BG} model.}
	\label{fig:diagram}
\end{figure}
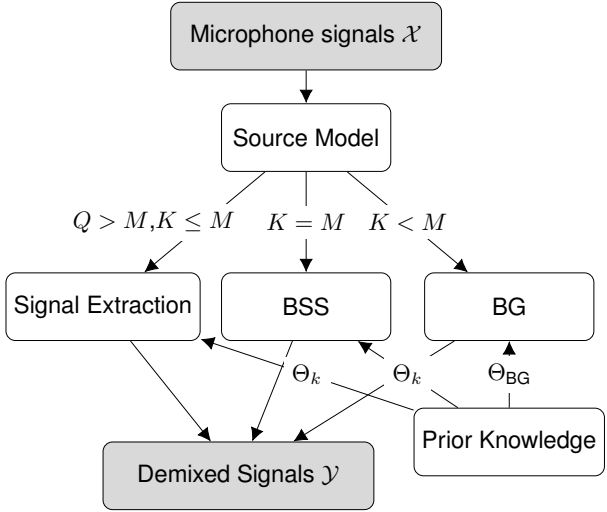
\Walter{Joining the priors for \acp{SOI} and \ac{BG}} yields the \Walter{overall} log prior term (neglecting constant terms) \Walter{(cf. \eqref{eq:prior_PDF})}
\begin{align}
&\log p(\SetW) = -\BlockIdxMax\sum_{\FreqIdx=1}^{\FreqIdxMax}\left(\gamma_\FreqIdx^\text{BG}\sum_{\ChannelIdx'=1}^{\MicIdxMax-\ChannelIdxMax}(\BGvecTransf_\FreqIdx^{\ChannelIdx'})\herm\PriorCovMatBG\BGvecTransf_\FreqIdx^{\ChannelIdx'}\dots\right.\label{eq:prior_costFctTerm}\\ 
&\left.\dots+\sum_{\ChannelIdx\in\SetIdxPrior}\gamma_{\ChannelIdx,\FreqIdx} (\w_\FreqIdx^\ChannelIdx)\herm\PriorCovMatInv\w_\FreqIdx^\ChannelIdx + \sum_{\ChannelIdx\in\mathcal{I}^\text{Euc}}\tilde{\gamma}_{\ChannelIdx,\FreqIdx}^{\text{Euc}}\Vert\w_\FreqIdx^\ChannelIdx - \SteeringVector_\FreqIdx(\vartheta_\ChannelIdx)\Vert_2^2\right),\notag
\end{align}
where we introduced the notation $\gamma_\FreqIdx^\text{BG} = \frac{\tilde{\gamma}_\FreqIdx^\text{BG}}{\BlockIdxMax}$, $\gamma_{\ChannelIdx,\FreqIdx} = \frac{\tilde{\gamma}_{\ChannelIdx,\FreqIdx}}{\BlockIdxMax}$ and $\gamma_{\ChannelIdx,\FreqIdx}^{\text{Euc}} = \frac{\tilde{\gamma}_{\ChannelIdx,\FreqIdx}^{\text{Euc}}}{\BlockIdxMax}$ for convenience in the following. The term contributing to the cost function is given by
\begin{equation}
	\BlockIdxMax J_\text{prior}(\SetW) = -\log p(\SetW).
	\label{eq:costFunction_prior}
\end{equation}
\subsection{Generic Cost Function}
\label{sec:generic_costFunction}
Taking the negative of the \ac{MAP} problem \eqref{eq:MAP_problem} and using \eqref{eq:BG_costFctTerm} and \eqref{eq:prior_costFctTerm} yields the generic cost function
\begin{align}
&J_\text{IBSS}(\SetW) = \underbrace{\sum_{\ChannelIdx=1}^{\ChannelIdxMax}\Expect{G\left(\sFreqVec_{\ChannelIdx,\BlockIdx}\right)} - 2\sum_{\FreqIdx=1}^{\FreqIdxMax}\log \left\vert \det \W_\FreqIdx\right\vert \dots}_{J_\text{BSS}(\SetW)} \notag\\
&\qquad\qquad\quad\dots + J_\text{BG}(\SetW) + J_\text{prior}(\SetW).
\label{eq:generic_cost_fct}
\end{align}
The cost function $J_\text{IBSS}$ consists of three parts: The \ac{BSS} cost function $J_\text{BSS}$, a component corresponding to the \ac{BG} $J_\text{BG}$ and a term representing the priors $J_\text{prior}$ of \acp{SOI} and \ac{BG}. Fig.~\ref{fig:diagram} gives an overview of different \Walter{tasks addressed by} the generic cost function \eqref{eq:generic_cost_fct}.

\subsection{Relation to \Walter{\ac{BSS}}}
By choosing \Walter{an uninformative} prior over the demixing matrices $p(\SetW) = \text{const.}$ and the number of \acp{SOI} equal to the number of microphones $\ChannelIdxMax = \MicIdxMax$, the cost function for non-informed \Walter{determined \ac{IVA}} is obtained \Walter{\cite{vincent_audio_2018}}
\begin{equation}
J_\text{BSS}(\SetW) = \sum_{\ChannelIdx=1}^{\ChannelIdxMax}\Expect{G\left(\sFreqVec_{\ChannelIdx,\BlockIdx}\right)} - 2\sum_{\FreqIdx=1}^{\FreqIdxMax}\log \left\vert \det \W_\FreqIdx\right\vert.
\label{eq:IVA_cost_function}
\end{equation}
\Walter{Hence, the proposed framework includes the prior work based on \ac{IVA} (and \ac{ICA} as a special case of \ac{IVA}) \WalterTwo{\cite{smaragdis_blind_1998,kim_blind_2007,ono_stable_2011,ono_auxiliary-function-based_2012} and its many extensions \cite{kitamura_determined_2016,scheibler_2019,mitsui_vectorwise_2018,brendel_2020}}.}
\section{Derivation of Update Rules}
\label{sec:MM_algorithm}
In the following, we develop an optimization algorithm based on the \ac{MM} principle for the \Walter{general} informed \ac{BSS} cost function $J_\text{IBSS}(\SetW)$ \eqref{eq:generic_cost_fct}. We will start with the fundamental \ac{MM} principle and \Walter{then construct} an upper bound of the informed \ac{BSS} cost function $J_\text{IBSS}$. Finally, we will provide update rules and summarize the proposed algorithmic framework.

\begin{figure}
	\setlength\fwidth{0.42\textwidth}
	\input{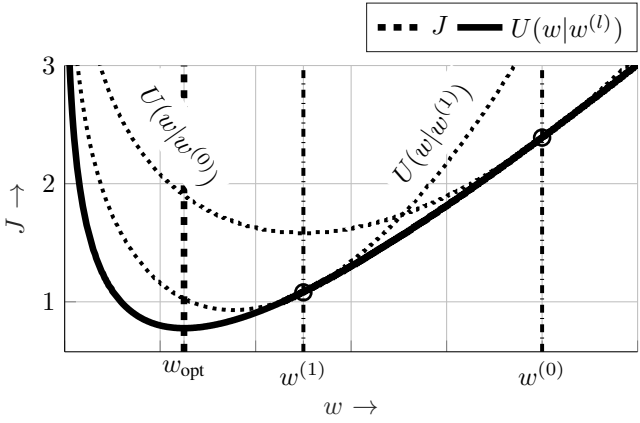}
	\caption{Illustration of optimization based on the \ac{MM} principle. Here, a one-dimensional cost function is used for illustration. The cost function $J$ is shown as a solid line and the upper bounds $\UpperBound(w \vert w^{(l)})$ for $\IterIdx = 0,1$ as dotted lines. Furthermore, the global minimizer $w_\text{opt}$ and the minimizer of $\UpperBound(w \vert w^{(0)})$ are shown as vertical lines.}
	\label{fig:illustration_MM}
\end{figure}
\subsection{Majorize-Minimize Principle}
The main idea of \Walter{\acf{MM}} algorithms is to define an upper bound \Walter{for the cost function} which is easier to optimize than the cost function \Walter{itself} and which fulfills two conditions: majorization and tangency \Walter{(see \cite{hunter_tutorial_2004} for an \WalterTwo{accessible in-depth} introduction)}. 

Let $\SetW^{(l)}$ denote the set of estimated demixing matrices at iteration $l\in[L]$ with \Walter{$L$ as the} \Walter{total} number of iterations. Then the majorization property of the upper bound $\UpperBound\left(\SetW\vert\SetW^{(l)}\right)$ can be expressed as
\begin{equation}
J(\SetW) \leq \UpperBound\left(\SetW\vert\SetW^{(l)}\right).
\end{equation}
Equality holds iff $\SetW = \SetW^{(l)}$, i.e.,
\begin{equation}
J\left(\SetW^{(l)}\right) = \UpperBound\left(\SetW^{(l)}\vert\SetW^{(l)}\right),
\end{equation}
which represents the tangency condition. The upper bound is chosen such that \Walter{its} optimization is easily possible 
\begin{equation}
\SetW^{(l+1)} =\underset{\SetW}{\text{argmin}} \ \UpperBound\left(\SetW\vert\SetW^{(l)}\right),
\label{eq:upperbound_minimization}
\end{equation}
where $\SetW^{(l+1)}$ denotes the minimizer. As minimization does not increase the function value of the upper bound, the following downhill property \cite{hunter_tutorial_2004} is obtained by using the tangency and majorization property of the upper bound
\begin{align}
J\left(\SetW^{(l+1)}\right) &\leq \UpperBound\left(\SetW^{(l+1)}\vert\SetW^{(l)}\right)\\
&\leq \UpperBound\left(\SetW^{(l)}\vert\SetW^{(l)}\right)=J\left(\SetW^{(l)}\right).\notag
\end{align}
Hence, by iteratively optimizing the upper bound and ensuring tangency to the cost function, the cost function values are ensured to be non-increasing.

This optimization principle is illustrated in Fig.~\ref{fig:illustration_MM}.
\subsection{Construction of Upper Bound}
The problem of optimizing the informed \ac{BSS} cost function $J_\text{IBSS}$ will now be shifted to optimizing a surrogate, an upper bound $\UpperBound_\text{IBSS}$.

Let $\SetW_\ChannelIdx^{(\IterIdx)} = \left\lbrace\w_\FreqIdx^{\ChannelIdx,(\IterIdx)} \in \mathbb{C}^{\ChannelIdxMax}\vert \FreqIdx \in \SetK\right\rbrace$ be the set of all demixing vectors for channel $\ChannelIdx$ at iteration $\IterIdx$. For supergaussian \acp{PDF} \Walter{(for the discussion of the time-varying Gaussian \ac{PDF} see below)}, characterized by the score function $G(\sFreqVec_{\ChannelIdx,\BlockIdx})$, the following inequality has been proven \Walter{in} \cite{ono_stable_2011}
\begin{equation}
\Expect{G(\sFreqVec_{\ChannelIdx,\BlockIdx})} \leq  R_\ChannelIdx(\SetW_\ChannelIdx^{(l)}) + \frac{1}{2}\sum_{\FreqIdx=1}^{\FreqIdxMax}\left(\w_\FreqIdx^\ChannelIdx\right)\herm \V_\FreqIdx^\ChannelIdx\left(\SetW_\ChannelIdx^{(l)}\right)\w_\FreqIdx^\ChannelIdx.
\label{eq:ono_inequality}
\end{equation}
\Walter{All discussed \ac{SOI} models can be written solely in dependence of the norm of the broadband \ac{SOI} signal $r_{\ChannelIdx,\FreqIdx,\BlockIdx}\Walter{(\SetW_\ChannelIdx^{(l)})}$, i.e., $\tilde{G}(r_{\ChannelIdx,\FreqIdx,\BlockIdx}\Walter{(\SetW_\ChannelIdx^{(l)})}) = G(\sFreqVec_{\ChannelIdx,\BlockIdx})$.} For the supergaussian and the time-varying Gaussian \ac{SOI} model, the weighting factor depends on the estimated broadband signal energy of source $\ChannelIdx$ at time instant $\BlockIdx$
\begin{equation}
r_{\ChannelIdx,\BlockIdx}\left(\SetW_\ChannelIdx^{(l)}\right) = \left\Vert \sFreqVec_{\ChannelIdx,\BlockIdx}^{(l)} \right\Vert_2 = \sqrt{\sum_{\FreqIdx = 1}^{\FreqIdxMax} \left\vert \left(\w_\FreqIdx^{\ChannelIdx,(l)}\right)\herm\x_{\FreqIdx,\BlockIdx} \right\vert^2},
\label{eq:signal_energy}
\end{equation}
i.e., $r_{\ChannelIdx,\FreqIdx,\BlockIdx} = r_{\ChannelIdx,\BlockIdx} \quad\forall \FreqIdx$. The term $R_\ChannelIdx(\SetW_\ChannelIdx^{(l)})$ \Walter{in \eqref{eq:ono_inequality}} given as
\begin{align}
&R_\ChannelIdx\left(\SetW_\ChannelIdx^{(l)}\right) = \hat{\mathbb{E}}\Bigg\lbrace \tilde{G}\left(r_{\ChannelIdx,\BlockIdx,\FreqIdx}\left(\SetW_\ChannelIdx^{(l)}\right)\right) \dots \\
&\qquad\qquad\qquad\dots-\frac{r_{\ChannelIdx,\BlockIdx,\FreqIdx}\left(\SetW_\ChannelIdx^{(l)}\right)\tilde{G}'\left(r_{\ChannelIdx,\BlockIdx,\FreqIdx}\left(\SetW_\ChannelIdx^{(l)}\right)\right)}{2}\Bigg\rbrace \notag
\end{align}
\Walter{is independent of $\SetW$ and} $\V_\FreqIdx^\ChannelIdx\left(\SetW_\ChannelIdx^{(l)}\right)$ denotes the weighted \Walter{sensor \WalterTwo{signals'}} covariance matrix
\begin{equation}
\V_\FreqIdx^\ChannelIdx\left(\SetW_\ChannelIdx^{(l)}\right) = \Expect{\phi(r_{\ChannelIdx,\FreqIdx,\BlockIdx})\x_{\FreqIdx,\BlockIdx}\x_{\FreqIdx,\BlockIdx}\herm}, 
\label{eq:weighted_covMat}
\end{equation}
where
\begin{equation}
\phi(r_{\ChannelIdx,\FreqIdx,\BlockIdx}) = \frac{\tilde{G}'\left(r_{\ChannelIdx,\FreqIdx,\BlockIdx}\left(\SetW_\ChannelIdx^{(l)}\right)\right)}{r_{\ChannelIdx,\FreqIdx,\BlockIdx}\left(\SetW_\ChannelIdx^{(l)}\right)}
\end{equation}
denotes the corresponding weighting factor. 

The weighting factor $\phi(r_{\ChannelIdx,\BlockIdx})$ for the generalized Gaussian distribution \eqref{eq:GGD_PDF} and the time-varying Gaussian \ac{PDF} \eqref{eq:SOI_model_timevaryingGaussian} can be expressed as \WalterTwo{(see \cite{ono_auxiliary-function-based_2012})}
\begin{equation}
	\phi(r_{\ChannelIdx,\BlockIdx}) = \left(r_{\ChannelIdx,\BlockIdx}\right)^{\beta-2}.
	\label{eq:weightingFactor_IVA}
\end{equation} For the \ac{NMF} source model, we obtain for the weighting factor
\begin{equation}
	\phi(r_{\ChannelIdx,\FreqIdx,\BlockIdx}) = \frac{1}{\left(\sum_{\BasisIdx=1}^{\BasisIdxMax}t_{\ChannelIdx,\FreqIdx,\BasisIdx}v_{\ChannelIdx,\BasisIdx,\BlockIdx}\right)^\beta}
	\label{eq:weightingFactor_NMF}.
\end{equation}
Note that the weighting factor $\phi(r_{\ChannelIdx,\BlockIdx,\FreqIdx})$ is frequency-dependent in the case of the \ac{NMF} source model.

The inequality \eqref{eq:ono_inequality} transforms the optimization of a general nonlinear function dependent on all frequency bins into the optimization of \Walter{the sum of} quadratic functions, \Walter{each of which dependent only on one frequency bin.} \Walter{The} dependency between the frequency bins is solely expressed by the weighting $\phi(r_{\ChannelIdx,\BlockIdx})$ of the microphone correlation matrix in \eqref{eq:weighted_covMat}. 

By \Walter{inserting the inequality \eqref{eq:ono_inequality} into the \ac{BSS} cost function \eqref{eq:IVA_cost_function}}, we obtain the following upper bound for the \ac{BSS} cost function $J_\text{BSS}$
\begin{align}
&\UpperBound_\text{BSS}\left(\SetW\vert\SetW^{(l)}\right)  = \sum_{\FreqIdx=1}^{\FreqIdxMax}\Bigg[ \sum_{\ChannelIdx=1}^{\ChannelIdxMax}\Bigg(\frac{1}{2}\left(\w_\FreqIdx^\ChannelIdx\right)\herm \V_\FreqIdx^\ChannelIdx\left(\SetW_\ChannelIdx^{(l)}\right)\w_\FreqIdx^\ChannelIdx \dots\notag \\
&\qquad \dots +  \frac{1}{\FreqIdxMax} R_\ChannelIdx\left(\SetW_\ChannelIdx^{(l)}\right)\Bigg) - 2\log \vert \det \W_\FreqIdx \vert\Bigg],\label{eq:upperBound_BSS}
\end{align}
with $J_\text{BSS}(\SetW) = \UpperBound_\text{BSS}\left(\SetW\vert\SetW^{(l)}\right)$ iff $\SetW = \SetW^{(l)}$.

For the case of a Gaussian source distribution, the upper bound is identical to the cost function (a similar relation holds for the \ac{NMF} source model described in Sec.~\ref{sec:NMF_sourceModel})
\begin{equation}
J_\text{BSS}\left(\SetW\vert\SetW^{(l)}\right)  = \UpperBound_\text{BSS}\left(\SetW\vert\SetW^{(l)}\right),
\end{equation}
where $R_\ChannelIdx\left(\SetW_\ChannelIdx^{(l)}\right) = 0$.

An upper bound of the cost function for informed \ac{BSS} $J_\text{IBSS}(\SetW)$ can be obtained by adding the cost function of the prior $J_\text{prior}$ \Walter{\eqref{eq:costFunction_prior}} and the cost function of the \ac{BG} $J_\text{BG}$ \Walter{\eqref{eq:BG_costFctTerm}} on both sides of the inequality
\begin{align}
	J_\text{IBSS}(\SetW) &\leq \UpperBound_\text{IBSS}\left(\SetW\vert\SetW^{(l)}\right)\label{eq:GBSS_upperBound}\\
	&= \UpperBound_\text{BSS}\left(\SetW\vert\SetW^{(l)}\right)+J_\text{BG}(\SetW)+J_\text{prior}(\SetW),\notag
\end{align}
with $J_\text{IBSS}(\SetW) = \UpperBound_\text{IBSS}\left(\SetW\vert\SetW^{(l)}\right)$ iff $\SetW = \SetW^{(l)}$, i.e., the upper bound fulfills the requirements of majorization and tangency.
\subsection{Optimization of Upper Bound}
In the following we will derive analytic expressions for the minimum of the upper bound w.r.t. the demixing matrices
\begin{equation}
\SetW^{(l+1)} =\underset{\SetW}{\text{argmin}} \ \UpperBound_\text{IBSS}\left(\SetW\vert\SetW^{(l)}\right)
\label{eq:upperbound_minimization_IBSS}
\end{equation}
and derive iterative update rules which allow the computation of the minimizer $\SetW^{(l+1)}$. To simplify the following derivation, we transform the log-det term of the upper bound \eqref{eq:upperBound_BSS} to have all \ac{BG} filters in the transformed representation \Walter{\eqref{eq:BG_transformation}}
\begin{align}
	\log \vert \det \W_\FreqIdx \vert &= \log \left\vert \det \begin{bmatrix}
	\mathbf{I}_{\ChannelIdxMax}&\mathbf{0}_{\MicIdxMax-\ChannelIdxMax\times\ChannelIdxMax}\\
	\mathbf{0}_{\ChannelIdxMax\times\MicIdxMax-\ChannelIdxMax}&\EigMat\ScaleMat\inv
	\end{bmatrix} \begin{bmatrix}
	\W_\FreqIdx^{\text{SOI}}\\
	\BGmatTransf_\FreqIdx
	\end{bmatrix} \right\vert\notag\\
	&= \log \left\vert \det \begin{bmatrix}
	\W_\FreqIdx^{\text{SOI}}\\
	\BGmatTransf_\FreqIdx
	\end{bmatrix} \right\vert + \text{const.}
\end{align} 
\Walter{Hence, the transformed filters yield the same optimum as the orignal filters.}
\subsubsection{Without Constraints}
For the unconstrained channels, i.e., for $\ChannelIdx\notin \mathcal{I}$ and $\ChannelIdx\notin \mathcal{I}^\text{Euc}$, \WalterTwo{we obtain the following conditions by setting the}
derivative of the upper bound \Walter{\eqref{eq:GBSS_upperBound}} w.r.t. \WalterTwo{each of} the \ac{SOI} filters to zero \cite{ono_stable_2011}
\begin{equation}
	\left(\w_\FreqIdx^q\right)\herm \V_\FreqIdx^\ChannelIdx\left(\SetW_\ChannelIdx^{(l)}\right)\w_\FreqIdx^\ChannelIdx \overset{!}{=}\delta_{\ChannelIdx q}\Walter{, \quad \ChannelIdx,q\in[\ChannelIdxMax]}
	\label{eq:SOI_quadrForm}
\end{equation}
where $\delta$ denotes the Kronecker Delta. Similarly, for the \ac{BG} filters we obtain \Walter{by differentiating \eqref{eq:GBSS_upperBound}} the following conditions for the relation between the \ac{SOI} filters $\ChannelIdx\in [\ChannelIdxMax]$ and the \ac{BG} filters $\ChannelIdx'\in [\MicIdxMax-\ChannelIdxMax]$ 
\begin{equation}
\left(\w_\FreqIdx^k\right)\herm \MicCorrMat_\FreqIdx\BGvecTransf_\FreqIdx^{\ChannelIdx'} \overset{!}{=}0
\label{eq:quadraticCondition_FGBG}
\end{equation}
and for the relation between the \ac{BG} filters
\begin{equation}
\left(\BGvecTransf_\FreqIdx^{q}\right)\herm \MicCorrMat_\FreqIdx\BGvecTransf_\FreqIdx^{\ChannelIdx'} \overset{!}{=}\delta_{\ChannelIdx' q}\Walter{, \quad q\in[\MicIdxMax-\ChannelIdxMax].}
\label{eq:BG_conditions}
\end{equation}
\Walter{However, this condition is \WalterTwo{not investigated further} in the following, as the estimation of the \ac{BG} signals is not our aim.}
By collecting all the vector-wise constraints in \eqref{eq:quadraticCondition_FGBG}, we can write
\begin{equation}
	\W_\FreqIdx^{\text{SOI}} \MicCorrMat_\FreqIdx\BGmatTransf_\FreqIdx\herm \overset{!}{=}\mathbf{0}_{\ChannelIdxMax\times(\MicIdxMax-\ChannelIdxMax)}.
\end{equation}
Now, we insert $\BGmatTransf_\FreqIdx = \ScaleMat\EigMat\herm\BGmat_\FreqIdx$
\begin{equation}
\W_\FreqIdx^{\text{SOI}} \MicCorrMat_\FreqIdx\BGmat_\FreqIdx\herm \EigMat\ScaleMat \overset{!}{=}\mathbf{0}_{\ChannelIdxMax\times(\MicIdxMax-\ChannelIdxMax)}
\end{equation}
and multiply with $\ScaleMat\inv\EigMat\herm$ from the right, which yields the following condition between \ac{SOI} and \ac{BG} filters
\begin{equation}
\W_\FreqIdx^{\text{SOI}} \MicCorrMat_\FreqIdx\BGmat_\FreqIdx\herm \overset{!}{=}\mathbf{0}_{\ChannelIdxMax\times(\MicIdxMax-\ChannelIdxMax)}.
\label{eq:condition_BGFG}
\end{equation}
\subsubsection{With Constraints}
For the channels constrained by \Walter{the quadratic constraint} \eqref{eq:directional_prior}, i.e., \mbox{$\ChannelIdx \in \mathcal{I}$}, we obtain as conditions for the \ac{SOI} channels \Walter{by optimizing \eqref{eq:GBSS_upperBound}}
\begin{equation}
	\left(\w_\FreqIdx^q\right)\herm \left[\V_\FreqIdx^\ChannelIdx\left(\SetW_\ChannelIdx^{(l)}\right) + \gamma_{\ChannelIdx,\FreqIdx}\PriorCovMatInv\right]\w_\FreqIdx^\ChannelIdx \overset{!}{=}\delta_{\ChannelIdx q}.
	\label{eq:SOI_quadrForm_constr}
\end{equation}
For the relation between the \ac{SOI} and the \ac{BG} channels we obtain
\begin{equation}
\W_\FreqIdx^{\text{SOI}} \left[\MicCorrMat_\FreqIdx+ \gamma_\FreqIdx^\text{BG}\PriorCovMatBG\right]\BGmat_\FreqIdx\herm \overset{!}{=}\mathbf{0}_{\ChannelIdxMax\times(\MicIdxMax-\ChannelIdxMax)}.
\label{eq:SOI_BG_quadrForm_constr}
\end{equation}
\WalterTwo{Note that the conditions \eqref{eq:SOI_quadrForm_constr} and \eqref{eq:SOI_BG_quadrForm_constr} \WalterFour{generalize the previously known conditions} \eqref{eq:SOI_quadrForm} and \eqref{eq:BG_conditions} \WalterFour{in the sense that} the weighted correlation matrix $\V_\FreqIdx^\ChannelIdx$ and \WalterFour{the} microphone signal correlation matrix $\MicCorrMat_\FreqIdx$ \WalterFour{are regularized} \WalterFour{by} the precision matrices \WalterFour{$\PriorCovMatInv$ and $\PriorCovMatBG$, which allow incorporation of many types of prior knowledge on \acp{SOI} and/or \ac{BG} as discussed in Sec.~\ref{sec:priors}}.}
\subsection{Update Rules}
\begin{table*}
	\centering
	\begin{tabular}{l|ccccccccccccc} 
		&\multicolumn{13}{c}{\Walter{Algorithm Index} $\rightarrow$}  \\  
		& 1              & 2              & 3 & 4 & 5 & 6 & 7 & 8 & 9 & 10 & 11 & 12 & 13\\ 
		\midrule
		$\ChannelIdxMax$        &$\MicIdxMax$            &$\MicIdxMax$& $\MicIdxMax$ &$\MicIdxMax$  &$1$  &$1$ &$1$ &$\MicIdxMax$  &$\MicIdxMax$  &$\MicIdxMax$  &$1$&$1$ &$1$ \\ 
		Optimization type       &GD&IP&IP&IP&IP&IP&IP&IP&IP&IP&IP&IP&IP\\
		Spatial One/Null       &One&One&One&Null&One&One&Null&One&One&Null&One&One&Null\\ 
	    Quadratic prior \eqref{eq:directional_prior} &\ding{53}&\ding{53}&\checkmark&\checkmark&\ding{53}&\checkmark&\checkmark&\ding{53}&\checkmark&\checkmark&\ding{53}&\checkmark&\checkmark\\ 
		Euclidean prior \eqref{eq:directional_prior_Eucl}      &\ding{53}&\checkmark&\ding{53}&\ding{53}&\checkmark&\ding{53}&\ding{53}&\checkmark&\ding{53}&\ding{53}&\checkmark&\ding{53}&\ding{53}\\ 
		\ac{BG} model                 &\ding{53}            &\ding{53}  &\ding{53}  &\ding{53}  &\checkmark  &\checkmark  &\checkmark  &\ding{53}  &\ding{53}  &\ding{53}  &\checkmark &\checkmark & \checkmark \\ 
		\ac{BG} prior           &\ding{53}    &\ding{53}  &\ding{53}  &\ding{53}  &\ding{53} &\ding{53}   &\checkmark  &\ding{53}  &\ding{53}  &\ding{53}  &\ding{53} &\ding{53} &\checkmark \\ 
		\ac{SOI} model   &   SG  &\multicolumn{6}{c}{----------------------- SG/TVG ---------------------}  &\multicolumn{6}{c}{------------------------ NMF ---------------------}\\
		Proposed                & \cite{vincent_geometrically_2015}&\multicolumn{2}{c}{--- New ---} &\cite{brendel_2020}&\multicolumn{3}{c}{---------- New -----------}&\cite{mitsui_vectorwise_2018}&\multicolumn{5}{c}{------------------ New -------------------}\\ 
		\bottomrule 
	\end{tabular}\vspace{1pt}
	\caption{Overview over algorithmic variants \Walter{evaluated} in the experiments. We used the following abbreviations: Gradient descent (GD), Iterative Projection (IP), Supergaussian (SG) and Time-Varying Gaussian (TVG).}
	\label{tab:algorithmic_variants}
\end{table*}
In the following, we will present update rules which \Walter{identify solutions to} the conditions \eqref{eq:SOI_quadrForm}, \eqref{eq:condition_BGFG}, \eqref{eq:SOI_quadrForm_constr} and \eqref{eq:SOI_BG_quadrForm_constr} presented in the previous paragraph.
\subsubsection{Demixing Filters}
In the unconstrained case the \ac{SOI} filters can be optimized by ensuring orthogonality between the output signals \cite{ono_stable_2011}
\begin{equation}
\tilde{\w}_\FreqIdx^{\ChannelIdx,(l+1)} = \left(\W_\FreqIdx^{\ChannelIdx,(l)}\V_\FreqIdx^{\ChannelIdx,(l)}\left(\SetW_\ChannelIdx^{(l)}\right)\right)^{-1}\mathbf{e}_\ChannelIdx,
\label{eq:update}
\end{equation}
where $\mathbf{e}_\ChannelIdx$ denotes a canonical basis vector with a one at the $\ChannelIdx$th position, and normalization
\begin{equation}
\w_\FreqIdx^{\ChannelIdx,(l+1)} = \frac{\tilde{\w}_\FreqIdx^{\ChannelIdx,(l+1)}}{\sqrt{\left(\tilde{\w}_\FreqIdx^{\ChannelIdx,(l+1)}\right)\herm \V_\FreqIdx^{\ChannelIdx,(l)}\left(\SetW_\ChannelIdx^{(l)}\right)\tilde{\w}_\FreqIdx^{\ChannelIdx,(l+1)}}}.
\label{eq:normalization}
\end{equation}
This procedure is called \ac{IP} and will be used to derive \Walter{generalized} update rules for the other algorithmic variants in the following. The channels constrained by \eqref{eq:directional_prior}, i.e., $\ChannelIdx\in \mathcal{I}$ are updated by
\begin{equation}
\tilde{\w}_\FreqIdx^{\ChannelIdx,(l+1)} = \left(\W_\FreqIdx^{(l)}\left[\V_\FreqIdx^{\ChannelIdx,(l)}\left(\SetW_\ChannelIdx^{(l)}\right) + \gamma_{\ChannelIdx,\FreqIdx}\PriorCovMatInv\right]\right)^{-1}\mathbf{e}_\ChannelIdx,
\label{eq:update_inf}
\end{equation}
\begin{equation}
\w_\FreqIdx^{\ChannelIdx,(l+1)} = \frac{\tilde{\w}_\FreqIdx^{\ChannelIdx,(l+1)}}{\sqrt{\left(\tilde{\w}_\FreqIdx^{\ChannelIdx,(l+1)}\right)\herm \left[\V_\FreqIdx^{\ChannelIdx,(l)}\left(\SetW_\ChannelIdx^{(l)}\right) + \gamma_{\ChannelIdx,\FreqIdx}\PriorCovMatInv\right]\tilde{\w}_\FreqIdx^{\ChannelIdx,(l+1)}}}.
\label{eq:normalization_inf}
\end{equation}
For the channels constrained by \eqref{eq:directional_prior_Eucl}, i.e., $\ChannelIdx\in \mathcal{I}^\text{Euc}$, we use the update rules proposed by \cite{mitsui_vectorwise_2018}
\begin{align}
	\mathbf{u}_\FreqIdx^\ChannelIdx &= \left(\W_\FreqIdx^{(\IterIdx)}\tilde{\V}_\FreqIdx^{\ChannelIdx,(\IterIdx)}\right)\inv\mathbf{e}_\ChannelIdx \label{eq:mitsui_updateFirst}\\
	\tilde{\mathbf{u}}_\FreqIdx^\ChannelIdx &= \gamma_{\ChannelIdx,\FreqIdx}^{\text{Euc}}\left(\tilde{\V}_\FreqIdx^{\ChannelIdx,(\IterIdx)}\right)\inv\SteeringVector_\FreqIdx(\vartheta_\ChannelIdx)\\
	p_{\ChannelIdx,\FreqIdx} &= (\mathbf{u}_\FreqIdx^\ChannelIdx)\herm\tilde{\V}_\FreqIdx^{\ChannelIdx,(\IterIdx)}\mathbf{u}_\FreqIdx^\ChannelIdx\\
	\tilde{p}_{\ChannelIdx,\FreqIdx} &= (\mathbf{u}_\FreqIdx^\ChannelIdx)\herm\tilde{\V}_\FreqIdx^{\ChannelIdx,(\IterIdx)}\tilde{\mathbf{u}}_\FreqIdx^\ChannelIdx\\
	\tilde{\w}_\FreqIdx^{\ChannelIdx,(l+1)} &\leftarrow \left\lbrace \begin{matrix}
	\frac{\mathbf{u}_\FreqIdx^\ChannelIdx}{\sqrt{p_{\ChannelIdx,\FreqIdx}}} + \tilde{\mathbf{u}}_\FreqIdx^\ChannelIdx, \qquad\quad\qquad\qquad \text{if} \quad \tilde{p}_{\ChannelIdx,\FreqIdx} = 0 \\
	\frac{\tilde{p}_{\ChannelIdx,\FreqIdx}}{2 p_{\ChannelIdx,\FreqIdx}}\left(-1+\sqrt{1+\frac{4 p_{\ChannelIdx,\FreqIdx}}{\vert \tilde{p}_{\ChannelIdx,\FreqIdx}\vert^2}}\right)\mathbf{u}_\FreqIdx^\ChannelIdx+\tilde{\mathbf{u}}_\FreqIdx^\ChannelIdx, \quad \text{else}.
	\end{matrix}\right.\label{eq:mitsui_updateLast}
\end{align}
To calculate the update of the \ac{BG} filters $\BGfilter_\FreqIdx$ in the unconstrained case, \eqref{eq:condition_BGFG} can be solved for $\BGfilter_\FreqIdx$ \Walter{by inserting the parametrization of the \ac{BG} filters}, which yields
\begin{equation}
\BGfilter_\FreqIdx = \left(\mathbf{E}_2 \mathbf{C}_\FreqIdx(\W_\FreqIdx^\text{SOI})\herm\right)\left(\mathbf{E}_1 \mathbf{C}_\FreqIdx(\W_\FreqIdx^\text{SOI})\herm\right)\inv.
\label{eq:BG_update}
\end{equation}
Hereby, we defined
\begin{equation}
\mathbf{E}_1 = [\mathbf{I}_\ChannelIdxMax,\mathbf{0}_{\ChannelIdxMax\times\MicIdxMax-\ChannelIdxMax}] \quad \text{and} \quad \mathbf{E}_2 = [\mathbf{0}_{\MicIdxMax-\ChannelIdxMax\times\ChannelIdxMax},\mathbf{I}_{\MicIdxMax-\ChannelIdxMax}].
\end{equation}
Note that these update rules coincide with those proposed by \cite{scheibler_2019}, but are rigorously derived here from the iterative projection perspective, which also makes the incorporation of priors possible.
Similarly, the updates for the constrained case are obtained \WalterFour{by generalization of \eqref{eq:BG_update}} as
\begin{align}
\BGfilter_\FreqIdx &= \left(\mathbf{E}_2 \left[\MicCorrMat_\FreqIdx+ \gamma_\FreqIdx^\text{BG}\PriorCovMatBG\right](\W_\FreqIdx^\text{SOI})\herm\right)\cdots \notag\\
&\quad \cdots\left(\mathbf{E}_1 \left[\MicCorrMat_\FreqIdx+ \gamma_\FreqIdx^\text{BG}\PriorCovMatBG\right](\W_\FreqIdx^\text{SOI})\herm\right)\inv.
\label{eq:BG_update_inf}
\end{align}
\subsubsection{Update of Demixed Signal Variance}
The update of the variance parameter $r_{\ChannelIdx,\BlockIdx,\FreqIdx}$ can be done directly based on the demixed signals for each iteration in case of the generalized Gaussian or time-varying Gaussian source model by \eqref{eq:signal_energy}. For the \ac{NMF} source model, the elements $t_{\ChannelIdx,\FreqIdx,\BasisIdx}$ of the basis vectors and the elements $v_{\ChannelIdx,\BasisIdx,\BlockIdx}$ of the activation vector have to be updated in addition to the demixing filters. The update rules are given \Walter{by} \cite{kitamura_determined_2016}
\begin{equation}
t_{\ChannelIdx,\FreqIdx,\BasisIdx} \leftarrow t_{\ChannelIdx,\FreqIdx,\BasisIdx}\sqrt{\frac{\sum_{\BlockIdx\in\SetN} \vert y_{\ChannelIdx,\FreqIdx,\BlockIdx}\vert^2 v_{\ChannelIdx,\BasisIdx,\BlockIdx}\left(r_{\BlockIdx,\FreqIdx}^\ChannelIdx\right)^{-2}}{\sum_{\BlockIdx\in\SetN} v_{\ChannelIdx,\BasisIdx,\BlockIdx} \left(r_{\BlockIdx,\FreqIdx}^\ChannelIdx\right)^{-1}}}
\end{equation}
and
\begin{equation}
v_{\ChannelIdx,\BasisIdx,\BlockIdx} \leftarrow v_{\ChannelIdx,\BasisIdx,\BlockIdx} \sqrt{\frac{\sum_{\FreqIdx\in\SetK} \vert y_{\ChannelIdx,\FreqIdx,\BlockIdx}\vert^2 t_{\ChannelIdx,\FreqIdx,\BasisIdx}\left(r_{\BlockIdx,\FreqIdx}^\ChannelIdx\right)^{-2}}{\sum_{\FreqIdx\in\SetK} t_{\ChannelIdx,\FreqIdx,\BasisIdx} \left(r_{\BlockIdx,\FreqIdx}^\ChannelIdx\right)^{-1}}}.
\end{equation}
\subsection{Practical Aspects}
In this paragraph, we discuss some aspects which are relevant for a practical realization of the \Walter{above algorithmic variants}. To avoid distortion of the signals by the scaling ambiguity in each frequency bin, the minimal distortion principle can be applied \cite{matsuoka_minimal_2002}. To avoid numerical instability of the algorithmic variants relying on an \ac{NMF} \ac{SOI} model, \cite{kitamura_effective_2017} proposed to normalize all estimated quantities in each iteration (see \cite{kitamura_effective_2017} for details).
The proposed algorithmic framework is summarized in Alg.~\ref{alg:pseudocode}.
\section{Experiments}
\label{sec:experiments}
\begin{table*}
	\centering
	\begin{tabular}{l|cccl|cccccccccccc} 
		&\multicolumn{16}{c}{\Walter{Algorithm Index} $\rightarrow$}  \\  
		             & 1    & \quad& \quad&            & 2 & 3 & 4 & 5 & 6 & 7 & 8 & 9 & 10 & 11 & 12 & 13\\ 
		\cmidrule{1-2} \cmidrule{4-17}
		Step size    &$0.05$& &&\Walter{$\gamma$,$\gamma^\text{Euc}$,$\gamma^\text{BG}$}     &$0.5$&$1.5$&$0.5$&$2$&$2$&$50$&$5$&$3$&$5$&$2.5$&$2.5$&$100$\\ 
		Prior Weight &$0.01$& && $\lambda_\text{Tik}$&$1$&$1$&$10^{-3}$&$1$&$1$&$10^{-3}$&$1$&$1$&$10^{-3}$&$1$&$1$&$10^{-3}$\\
		             &      && &\Walter{$\lambda_1^\text{One}$,$\lambda_1^\text{Zero}$}   &\ding{53}&$2$&$1$&\ding{53}&$1.5$&$1$&\ding{53}&$1.5$&$1$&\ding{53}&$1$ &$1$\\ 
		             &      && &$\BasisIdxMax$&\ding{53}&\ding{53}&\ding{53}&\ding{53}&\ding{53}&\ding{53}&$2$&$2$&$2$&$2$&$2$&$2$\\
		$\IterIdxMax$&$2500$ & &&$\IterIdxMax$ &$100$ &$100$ &$100$ &$100$ &$100$ &$100$ &$100$ &$100$ &$100$ &$100$ &$100$ &$100$\\ 
		\cmidrule{1-2} \cmidrule{4-17}
	\end{tabular}\vspace{1pt}
	\caption{Parameters used in the experiments.}\vspace{-5pt}
	\label{tab:parameters}
\end{table*}
\begin{figure}
	\flushright
	\setlength\fwidth{0.42\textwidth}
%
%
\begin{tikzpicture}

\begin{axis}[%
width=1\fwidth,
height=0.55\fwidth,
at={(0\fwidth,0\fwidth)},
scale only axis,
xmin=1,
xmax=13,
xtick={1,2,3,4,5,6,7,8,9,10,11,12,13},
xticklabels={,,,,,,,,,,,,},
xlabel style={font=\color{white!15!black}},
ymin=-5,
ymax=10,
ylabel style={font=\color{white!15!black}},
ylabel={$\mathrm{dB}$ $\rightarrow$},
axis background/.style={fill=white},
xmajorgrids,
ymajorgrids,
legend style={at = {(1,1.3)}, legend cell align=left, align=left, draw=white!15!black}, legend columns = 3
]

\node[fill = white] at (axis cs: 2.2,8.3) {\boxed{\textbf{$\Delta$SDR}}};

\addplot [color=black, mark=x, mark options={solid, black}, mark size = 3, line width = 1pt]
table[row sep=crcr]{%
	1	4.1992\\
	2	3.7917\\
	3	2.4862\\
	4	4.273\\
	5	2.5833\\
	6	2.2364\\
	7	5.365\\
	8	8.4206\\
	9	8.2712\\
	10	3.5486\\
	11	6.5244\\
	12	1.9693\\
	13	5.7734\\
};
\addlegendentry{Room 1 $\SrcIdx = 1$}

\addplot [color=black, mark=asterisk, mark options={solid, black}, mark size = 3, line width = 1pt]
table[row sep=crcr]{%
	1	1.6415\\
	2	2.1435\\
	3	1.1422\\
	4	3.026\\
	5	1.4403\\
	6	1.1959\\
	7	3.4009\\
	8	5.5907\\
	9	5.0752\\
	10	4.3313\\
	11	2.6041\\
	12	1.4288\\
	13	2.9759\\
};
\addlegendentry{Room 1 $\SrcIdx = 2$}

\addplot [color=black, mark=o, mark options={solid, black}, mark size = 3, line width = 1pt]
table[row sep=crcr]{%
	1	4.6078\\
	2	3.3025\\
	3	2.7423\\
	4	5.7461\\
	5	2.8553\\
	6	2.484\\
	7	5.9019\\
	8	7.9349\\
	9	7.6238\\
	10	5.1636\\
	11	8.5103\\
	12	3.9667\\
	13	9.9096\\
};
\addlegendentry{Room 1 $\SrcIdx = 3$}

\addplot [color=black, dashed, mark=x, mark options={solid, black}, mark size = 3, line width = 1pt]
table[row sep=crcr]{%
	1	4.2806\\
	2	2.9333\\
	3	1.6076\\
	4	-4.9693\\
	5	1.8391\\
	6	1.0318\\
	7	-3.8579\\
	8	5.0407\\
	9	5.1637\\
	10	-3.8584\\
	11	3.8721\\
	12	1.908\\
	13	-1.3247\\
};
\addlegendentry{Room 2 $\SrcIdx = 1$}

\addplot [color=black, dashed, mark=asterisk, mark options={solid, black}, mark size = 3, line width = 1pt]
table[row sep=crcr]{%
	1	2.808\\
	2	1.8024\\
	3	1.211\\
	4	-0.0561\\
	5	1.449\\
	6	1.2169\\
	7	1.928\\
	8	3.9986\\
	9	3.8997\\
	10	-0.3802\\
	11	2.3619\\
	12	1.7767\\
	13	1.7484\\
};
\addlegendentry{Room 2 $\SrcIdx = 2$}

\addplot [color=black, dashed, mark=o, mark options={solid, black}, mark size = 3, line width = 1pt]
table[row sep=crcr]{%
	1	4.692\\
	2	3.3737\\
	3	1.5264\\
	4	-2.2076\\
	5	2.6225\\
	6	2.1028\\
	7	-2.4408\\
	8	6.4724\\
	9	7.668\\
	10	-2.6064\\
	11	6.7965\\
	12	5.9483\\
	13	-1.1926\\
};
\addlegendentry{Room 2 $\SrcIdx = 3$}

\end{axis}
\end{tikzpicture}%
%
%
\begin{tikzpicture}

\begin{axis}[%
width=1\fwidth,
height=0.55\fwidth,
at={(0\fwidth,0\fwidth)},
scale only axis,
xmin=1,
xmax=13,
xtick={1,2,3,4,5,6,7,8,9,10,11,12,13},
xticklabels={,,,,,,,,,,,,},
xlabel style={font=\color{white!15!black}},
ymin=4.5,
ymax=25,
ylabel style={font=\color{white!15!black}},
ylabel={$\mathrm{dB}$ $\rightarrow$},
axis background/.style={fill=white},
xmajorgrids,
ymajorgrids,
legend style={legend cell align=left, align=left, draw=white!15!black}
]

\node[fill = white] at (axis cs: 2.1,23) {\boxed{\textbf{$\Delta$SIR}}};

\addplot [color=black, mark=x, mark options={solid, black}, mark size = 3, line width = 1pt]
table[row sep=crcr]{%
	1	11.269\\
	2	16.488\\
	3	13.993\\
	4	15.146\\
	5	13.027\\
	6	12.693\\
	7	18.827\\
	8	23.984\\
	9	23.846\\
	10	13.81\\
	11	20.481\\
	12	7.8552\\
	13	18.443\\
};

\addplot [color=black, mark=asterisk, mark options={solid, black}, mark size = 3, line width = 1pt]
table[row sep=crcr]{%
	1	8.8396\\
	2	13.484\\
	3	11.748\\
	4	12.428\\
	5	12.813\\
	6	12.285\\
	7	13.569\\
	8	20.01\\
	9	19.241\\
	10	14.763\\
	11	14.068\\
	12	10.828\\
	13	12.933\\
};

\addplot [color=black, mark=o, mark options={solid, black}, mark size = 3, line width = 1pt]
table[row sep=crcr]{%
	1	10.006\\
	2	12.477\\
	3	10.684\\
	4	13.845\\
	5	11.525\\
	6	10.843\\
	7	17.165\\
	8	19.022\\
	9	18.437\\
	10	13.816\\
	11	23.147\\
	12	11.167\\
	13	24.399\\
};

\addplot [color=black, dashed, mark=x, mark options={solid, black}, mark size = 3, line width = 1pt]
table[row sep=crcr]{%
	1	12.354\\
	2	17.295\\
	3	13.015\\
	4	5.8136\\
	5	14.316\\
	6	12.674\\
	7	10.2685\\
	8	20.443\\
	9	21.762\\
	10	4.7518\\
	11	19.14\\
	12	12.094\\
	13	11.3658\\
};

\addplot [color=black, dashed, mark=asterisk, mark options={solid, black}, mark size = 3, line width = 1pt]
table[row sep=crcr]{%
	1	9.1922\\
	2	12.407\\
	3	11.091\\
	4	8.742\\
	5	14.098\\
	6	13.536\\
	7	12.591\\
	8	16.105\\
	9	14.798\\
	10	6.683\\
	11	13.03\\
	12	9.015\\
	13	11.989\\
};

\addplot [color=black, dashed, mark=o, mark options={solid, black}, mark size = 3, line width = 1pt]
table[row sep=crcr]{%
	1	10.398\\
	2	13.847\\
	3	10.762\\
	4	6.2273\\
	5	11.913\\
	6	10.716\\
	7	8.9446\\
	8	17.335\\
	9	19.501\\
	10	6.41875\\
	11	21.617\\
	12	16.32\\
	13	10.0018\\
};

\end{axis}
\end{tikzpicture}%
%
%
\begin{tikzpicture}

\begin{axis}[%
width=1\fwidth,
height=0.55\fwidth,
at={(0\fwidth,0\fwidth)},
scale only axis,
xmin=1,
xmax=13,
xtick={1,2,3,4,5,6,7,8,9,10,11,12,13},
xticklabels={,,,,,,,,,,,,},
ymin=-2,
ymax=11,
ylabel style={font=\color{white!15!black}},
ylabel={$\mathrm{dB}$ $\rightarrow$},
axis background/.style={fill=white},
xmajorgrids,
ymajorgrids,
legend style={legend cell align=left, align=left, draw=white!15!black}
]

\node[fill = white] at (axis cs: 2.1,9.5) {\boxed{\textbf{$\Delta$SAR}}};

\addplot [color=black, mark=x, mark options={solid, black}, mark size = 3, line width = 1pt]
table[row sep=crcr]{%
	1	6.9623\\
	2	4.422\\
	3	3.4008\\
	4	5.3219\\
	5	3.8339\\
	6	3.6957\\
	7	5.955\\
	8	8.7547\\
	9	8.6471\\
	10	4.8099\\
	11	6.8298\\
	12	4.9303\\
	13	7.2994\\
};

\addplot [color=black, mark=asterisk, mark options={solid, black}, mark size = 3, line width = 1pt]
table[row sep=crcr]{%
	1	4.8188\\
	2	3.4154\\
	3	2.5841\\
	4	4.5101\\
	5	2.5377\\
	6	2.3304\\
	7	4.7115\\
	8	5.9606\\
	9	5.5875\\
	10	5.1634\\
	11	4.7578\\
	12	4.426\\
	13	4.0454\\
};

\addplot [color=black, mark=o, mark options={solid, black}, mark size = 3, line width = 1pt]
table[row sep=crcr]{%
	1	7.5207\\
	2	4.9933\\
	3	4.3768\\
	4	7.1701\\
	5	4.831\\
	6	4.4376\\
	7	6.7477\\
	8	8.4199\\
	9	8.6516\\
	10	6.9236\\
	11	8.7369\\
	12	6.7039\\
	13	10.304\\
};

\addplot [color=black, dashed, mark=x, mark options={solid, black}, mark size = 3, line width = 1pt]
table[row sep=crcr]{%
	1	5.0974\\
	2	3.3758\\
	3	2.3353\\
	4	0.5476\\
	5	2.6823\\
	6	2.2984\\
	7	-1.4859\\
	8	5.3507\\
	9	5.3798\\
	10	2.8552\\
	11	4.3775\\
	12	5.1898\\
	13	-0.4587\\
};

\addplot [color=black, dashed, mark=asterisk, mark options={solid, black}, mark size = 3, line width = 1pt]
table[row sep=crcr]{%
	1	4.6721\\
	2	2.5357\\
	3	2.3091\\
	4	2.3162\\
	5	2.0351\\
	6	1.9396\\
	7	3.0204\\
	8	4.4974\\
	9	4.681\\
	10	3.139\\
	11	3.6454\\
	12	3.692\\
	13	2.7422\\
};

\addplot [color=black, dashed, mark=o, mark options={solid, black}, mark size = 3, line width = 1pt]
table[row sep=crcr]{%
	1	7.6441\\
	2	4.7522\\
	3	3.5895\\
	4	1.7281\\
	5	4.6428\\
	6	4.2889\\
	7	1.72248\\
	8	7.3697\\
	9	8.0535\\
	10	4.5365\\
	11	7.1725\\
	12	6.8983\\
	13	2.074178\\
};

\end{axis}
\end{tikzpicture}%
	\hspace*{-6pt}\input{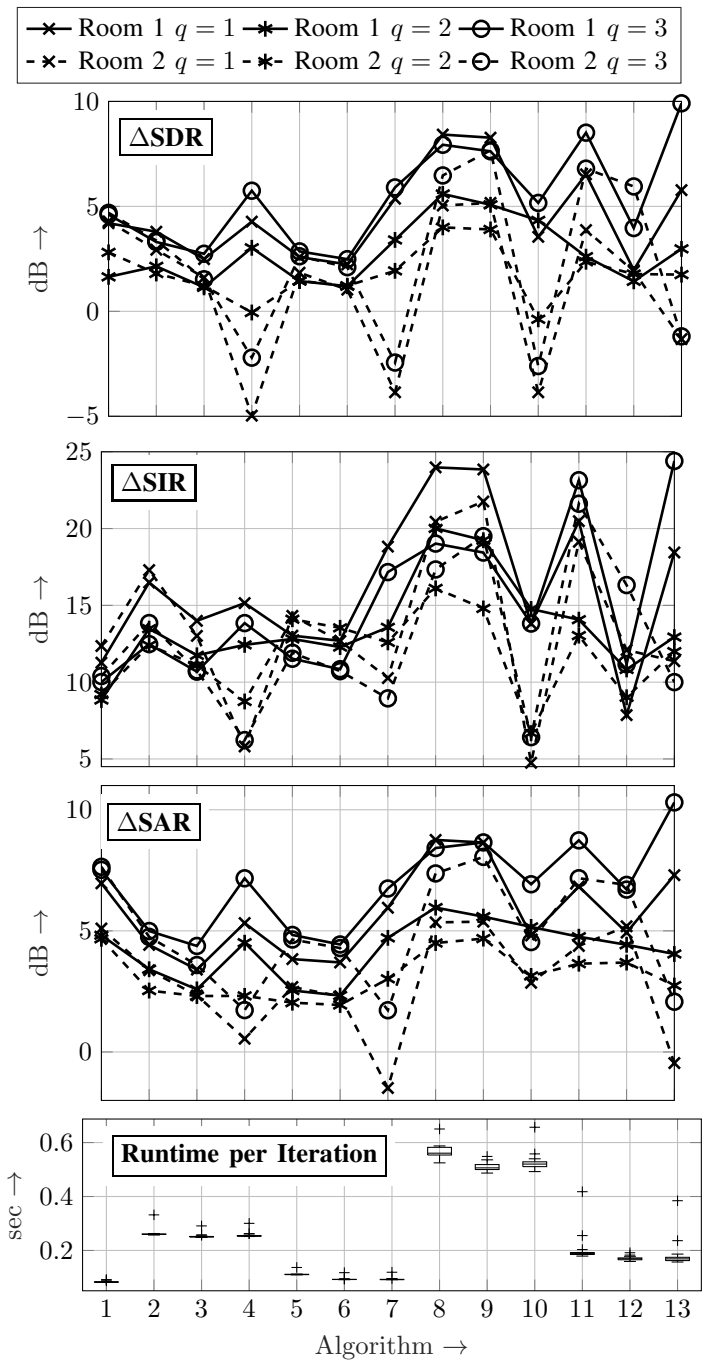}
	\caption{\Walter{Improvement in performance} measures \cite{vincent_performance_2006} \Walter{and average runtime per iteration} for different extracted sources ($\SrcIdx = 1,2,3$, see Fig.~\ref{fig:setup} for the geometric setup) and two different rooms: Room 1 with $T_{60} = 0.2\,\mathrm{s}$ and Room 2 with $T_{60} = 0.4\,\mathrm{s}$.}
	\label{fig:differentScenarios}
\end{figure}
\begin{algorithm}
	\caption{Informed \ac{BSS} (generic pseudo code)}
	\label{alg:pseudocode}
	\begin{algorithmic}
		\STATE \textbf{INPUT:} $\X$, $L$, $\left\lbrace\Theta_\ChannelIdx\right\rbrace_{\ChannelIdx\in\mathcal{I}}$, $\left\lbrace\Theta_\ChannelIdx\right\rbrace_{\ChannelIdx\in\mathcal{I}^\text{Euc}}$, $\Theta_\text{BG}$\vspace{-2pt}
		\STATE ------------------------------------------------------------------------\vspace{-2pt}
		\STATE \textbf{INITIALIZATION:} \STATE $\y_{\FreqIdx,\BlockIdx} = \x_{\FreqIdx,\BlockIdx}$ $\forall \FreqIdx,\BlockIdx$
		\IF{NMF Source Model}
		\STATE $t_{\ChannelIdx,\FreqIdx,\BasisIdx}, v_{\ChannelIdx,\BasisIdx,\BlockIdx} \sim \mathcal{U}(0,1)$ $\forall \ChannelIdx,\FreqIdx,\BlockIdx,\BasisIdx$
		\ENDIF
		\IF{$\MicIdxMax \leq \ChannelIdxMax$}
		\STATE $\W_\FreqIdx^{(0)} = \mathbf{I}_\MicIdxMax$ $\forall \FreqIdx$
		\ELSE
		\STATE $\W_\FreqIdx^{(0)} = \begin{bmatrix}
		\mathbf{I}_\ChannelIdxMax & \mathbf{0}_{\ChannelIdxMax\times(\MicIdxMax-\ChannelIdxMax)}\\
		\mathbf{0}_{(\MicIdxMax-\ChannelIdxMax)\times\ChannelIdxMax}&-\mathbf{I}_{\MicIdxMax-\ChannelIdxMax}
		\end{bmatrix}$ $\forall \FreqIdx$
		\ENDIF
		\STATE------------------------------------------------------------------------\vspace{-2pt}		
		\FOR{$l=1$ \TO $L$} 
		\FOR{$\ChannelIdx=1$ \TO $\ChannelIdxMax$} 
		\STATE Calculate $\phi(r_{\ChannelIdx,\FreqIdx,\BlockIdx})$
		$\forall \BlockIdx$ by \eqref{eq:weightingFactor_IVA} or \eqref{eq:weightingFactor_NMF}
		\FOR{$\FreqIdx=1$ \TO $\FreqIdxMax$}
		\STATE Calculate $\V_\FreqIdx^\ChannelIdx(\SetW_\ChannelIdx^{(l)}) = \Expect{\phi(r_{\ChannelIdx,\FreqIdx,\BlockIdx})\x_{\FreqIdx,\BlockIdx}\x_{\FreqIdx,\BlockIdx}\herm}$
		\IF{$\ChannelIdx\in\mathcal{I}$ or $\ChannelIdx\in\mathcal{I}^\text{Euc}$}
		\STATE Update $\w_\FreqIdx^\ChannelIdx$ by \eqref{eq:update_inf}, \eqref{eq:normalization_inf} or by \eqref{eq:mitsui_updateFirst}-\eqref{eq:mitsui_updateLast}
		\ELSIF{$\ChannelIdx\notin\mathcal{I}$}
		\STATE Update $\w_\FreqIdx^\ChannelIdx$ by \eqref{eq:update} and \eqref{eq:normalization}
		\ENDIF
		\IF{$\MicIdxMax > \ChannelIdxMax$}
		\IF{$\Theta_\text{BG}\neq\emptyset$}
		\STATE Update $\BGfilter_\FreqIdx$ by \eqref{eq:BG_update_inf}
		\ELSE
		\STATE Update $\BGfilter_\FreqIdx$ by \eqref{eq:BG_update}
		\ENDIF
		\ENDIF
		\STATE Assemble $\W_\FreqIdx = \begin{bmatrix}
		\begin{bmatrix}
		\w_\FreqIdx^1,\dots,\w_\FreqIdx^\ChannelIdxMax
		\end{bmatrix}\herm\\\begin{bmatrix}
		\BGfilter_\FreqIdx&-\mathbf{I}_{\MicIdxMax-\ChannelIdxMax}
		\end{bmatrix}
		\end{bmatrix}$
		\ENDFOR
		\ENDFOR
		\IF{NMF Source Model}
		\STATE Normalize \cite{kitamura_effective_2017}
		\ENDIF
		\ENDFOR
		\STATE Scale demixing filters $\W_\FreqIdx \leftarrow \text{diag}\left\lbrace\left(\W_\FreqIdx\right)\inv\right\rbrace\W_\FreqIdx$
		\FOR{$\BlockIdx=1$ \TO $\BlockIdxMax$}
		\FOR{$\FreqIdx=1$ \TO $\FreqIdxMax$}
		\STATE Extract \acp{SOI} $\s_{\FreqIdx,\BlockIdx} = \W_\FreqIdx^\text{SOI} \x_{\FreqIdx,\BlockIdx}$
		\ENDFOR
		\ENDFOR
		\STATE ------------------------------------------------------------------------\vspace{-2pt}
		\STATE \textbf{OUTPUT:} \acp{SOI} $\s_{\FreqIdx,\BlockIdx}\forall\FreqIdx,\BlockIdx$
	\end{algorithmic}
\end{algorithm}
In this section, we evaluate different algorithmic variants \Walter{resulting} from the proposed framework and compare them with several baseline algorithms from the literature. In this experimental study, we will focus on signal extraction, i.e., the separation from one source out of the observed mixture. \Walter{In addition}, the challenging case of an underdetermined scenario, i.e., $\SrcIdxMax > \MicIdxMax$ is \Walter{addressed} in the experiments in the following. However, also the extraction of multiple sources from the mixture and source separation for the determined case, i.e., $\ChannelIdxMax = \MicIdxMax$, and the overdetermined case, i.e.,  $\ChannelIdxMax > \MicIdxMax$, are \Walter{covered by} the framework. We do not evaluate the determined case here as this has been subject to many experimental studies in the literature \Walter{\cite{kim_blind_2007,ono_auxiliary-function-based_2012}}. We also do not investigate the overdetermined case, as this can be considered as an easier problem than the underdetermined scenario. A discussion for the overdetermined case \Walter{without the incorporation of prior knowledge} can be found in \cite{scheibler_2019}.
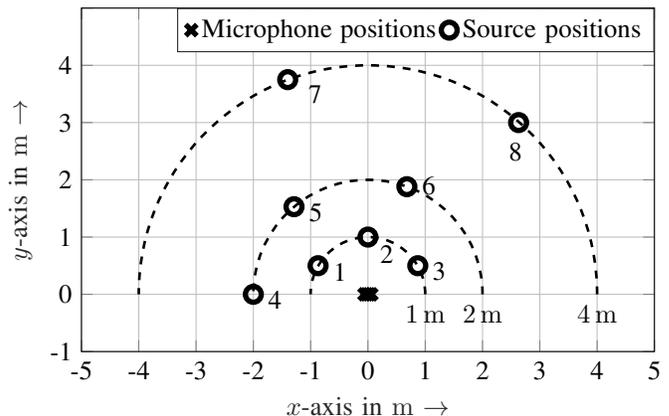
\begin{figure}
	\setlength\fwidth{0.42\textwidth}
%
%
%
\begin{tikzpicture}

\begin{axis}[%
width=1\fwidth,
height=0.6\fwidth,
at={(0\fwidth,0\fwidth)},
scale only axis,
xmin=-2,
xmax=8,
xlabel style={font=\color{white!15!black}},
xlabel={$x$-axis in $\mathrm{m}$ $\rightarrow$},
ymin=0,
ymax=6,
ylabel style={font=\color{white!15!black}},
ylabel={$y$-axis in $\mathrm{m}$ $\rightarrow$},
axis background/.style={fill=white},
xmajorgrids,
ymajorgrids,
legend style={at={(1,1)},legend cell align=left, align=left, draw=white!15!black}, legend columns = 2, xtick = {-2,-1,0,1,2,3,4,5,6,7,8}, ytick = {0,1,2,3,4,5,6,7}, xticklabels = {-5,-4,-3,-2,-1,0,1,2,3,4,5}, yticklabels = {-1,0,1,2,3,4}
]
\addlegendimage{only marks, color=black, draw=none, mark size=3.0pt, mark=x, mark options={solid, black}, line width = 2pt},
\addlegendimage{only marks, color=black, draw=none, mark size=3.0pt, mark=o, mark options={solid, black}, line width = 2pt},
\addplot [color=black, draw=none, mark size=3.0pt, mark=x, mark options={solid, black}, line width = 2pt]
  table[row sep=crcr]{%
2.94	1\\
2.98	1\\
3.02	1\\
3.06	1\\
};
\addlegendentry{Microphone positions}

\addplot [color=black, draw=none, mark size=3.0pt, mark=o, mark options={solid, black}, line width = 2pt]
  table[row sep=crcr]{%
2.13	1.5\\
3	2\\
3.87	1.5\\
1	1\\
5.63	4\\
1.6	4.75\\
3.68	2.88\\
1.71	2.53\\
};
\addlegendentry{Source positions}

\node[right, align=left, font=\color{black}]
at (axis cs:2.234,1.4) {1};
\node[right, align=left, font=\color{black}]
at (axis cs:3.05,1.7) {2};
\node[right, align=left, font=\color{black}]
at (axis cs:3.966,1.4) {3};
\node[right, align=left, font=\color{black}]
at (axis cs:1.1,0.9) {4};
\node[right, align=left, font=\color{black}]
at (axis cs:5.3,3.5) {8};
\node[right, align=left, font=\color{black}]
at (axis cs:1.8,4.5) {7};
\node[right, align=left, font=\color{black}]
at (axis cs:3.784,2.9) {6};
\node[right, align=left, font=\color{black}]
at (axis cs:1.814,2.432) {5};

\draw[dashed,line width = 1pt] (4,1)arc(0:180:1);
\draw[dashed,line width = 1pt] (5,1)arc(0:180:2);
\draw[dashed,line width = 1pt] (7,1)arc(0:180:4);

\node at (axis cs:4,0.65) {$1\, \mathrm{m}$};
\node at (axis cs:5,0.65) {$2\, \mathrm{m}$};
\node at (axis cs:7,0.65) {$4\, \mathrm{m}$};

\end{axis}
\end{tikzpicture}%
	\caption{Geometric setup of the scenario used in the experiments. The $\MicIdxMax = 4$ microphone positions are marked by crosses and the $Q = 8$ source positions at $1\,\mathrm{m}$, $2\,\mathrm{m}$ and $4\,\mathrm{m}$ distance from the array are marked by circles.}
	\label{fig:setup}
\end{figure}

The discussed methods \Walter{vary} w.r.t. the used \ac{SOI} model, the exploitation of a \ac{BG} model, the optimization method and the applied priors. Method 1 is based on gradient descent and a supergaussian source model and has been proposed in \cite{vincent_geometrically_2015}. The rest of the discussed algorithmic variants \Walter{all use \ac{IP} for optimization and} are evaluated for different \ac{SOI} models: the supergaussian, the time-varying Gaussian and the \ac{NMF} \ac{SOI} model.  For each of these \ac{SOI} models, we discuss the priors \eqref{eq:directional_prior} with \eqref{eq:precision_matrix_one} and \eqref{eq:directional_prior_Eucl} constraining one channel by a spatial one and the prior \eqref{eq:directional_prior} with \eqref{eq:precision_matrix_null} constraining all channels but one with a spatial null. Furthermore, we discuss for all source models the incorporation of the \ac{BG} model in two different variants: \Walter{1) unconstrained \ac{BG} \WalterTwo{with} a spatial one constraint for the \ac{SOI} (\eqref{eq:directional_prior} with \eqref{eq:precision_matrix_one} or \eqref{eq:directional_prior_Eucl}) and 2) unconstrained \ac{SOI}, but \ac{BG} with a \WalterTwo{spatial} null constraint \eqref{eq:directional_prior_BG}.} Tab.~\ref{tab:algorithmic_variants} summarizes the 13 algorithmic variants discussed in the following. The variants 4 and 8 are published in \cite{brendel_2020} and \cite{mitsui_vectorwise_2018}, respectively, and represent further baselines in our experimental study. Note that \cite{scheibler_2019}, which is a special case of the proposed framework, has been shown to be superior to \cite{koldovsky_gradient_2019} by comprehensive experiments. Hence, we do not repeat these experiments here.
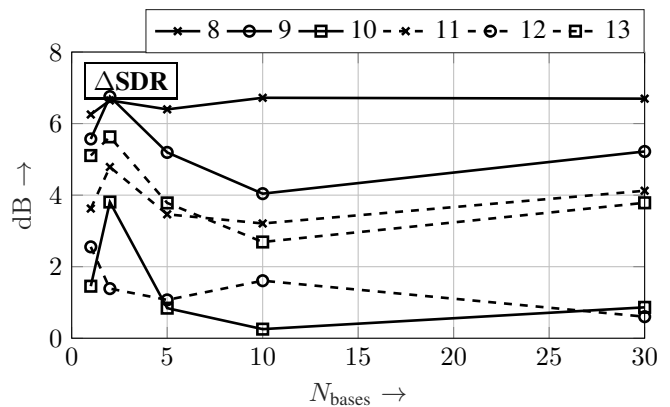
\begin{figure}
	\setlength\fwidth{0.42\textwidth}
%
%
%
\begin{tikzpicture}

\begin{axis}[%
width=1\fwidth,
height=0.5\fwidth,
at={(0\fwidth,0\fwidth)},
scale only axis,
xmin=0,
xmax=30,
xlabel style={font=\color{white!15!black}},
xlabel={$\BasisIdxMax$ $\rightarrow$},
ymin=0,
ymax=8,
ylabel style={font=\color{white!15!black}},
ylabel={$\mathrm{dB}$ $\rightarrow$},
axis background/.style={fill=white},
xmajorgrids,
ymajorgrids,
legend style={at = {(1,1.15)}, legend cell align=left, align=left, draw=white!15!black}, legend columns = 6
]

\node[fill = white] at (axis cs: 3.1,7.2) {\boxed{\textbf{$\Delta$SDR}}};

\addplot [color=black, mark=x, mark options={solid, black}, line width = 1pt]
table[row sep=crcr]{%
	1	6.2539\\
	2	6.6473\\
	5	6.3976\\
	10	6.721\\
	30	6.6967\\
};
\addlegendentry{8}

\addplot [color=black, mark=o, mark options={solid, black}, line width = 1pt]
table[row sep=crcr]{%
	1	5.5702\\
	2	6.7543\\
	5	5.1944\\
	10	4.0414\\
	30	5.2204\\
};
\addlegendentry{9}

\addplot [color=black, mark=square, mark options={solid, black}, line width = 1pt]
table[row sep=crcr]{%
	1	1.4549\\
	2	3.8076\\
	5	0.8414\\
	10	0.2548\\
	30	0.8639\\
};
\addlegendentry{10}

\addplot [color=black, dashed, mark=x, mark options={solid, black}, line width = 1pt]
table[row sep=crcr]{%
	1	3.6254\\
	2	4.7862\\
	5	3.464\\
	10	3.2083\\
	30	4.1227\\
};
\addlegendentry{11}

\addplot [color=black, dashed, mark=o, mark options={solid, black}, line width = 1pt]
table[row sep=crcr]{%
	1	2.5543\\
	2	1.3876\\
	5	1.072\\
	10	1.6082\\
	30	0.60292\\
};
\addlegendentry{12}

\addplot [color=black, dashed, mark=square, mark options={solid, black}, line width = 1pt]
table[row sep=crcr]{%
	1	5.1076\\
	2	5.6282\\
	5	3.7824\\
	10	2.6906\\
	30	3.7861\\
};
\addlegendentry{13}

\end{axis}

\end{tikzpicture}%
	\caption{Results of the number of bases $\BasisIdxMax$ for the algorithmic variants using an \ac{NMF} source model. The results for the approaches using a \ac{BG} model are depicted as solid lines, the others as dashed lines.}
	\label{fig:num_bases}
\end{figure}
\subsection{Experimental Setup}
For the experiments we used a uniform linear array with $\MicIdxMax = 4$ microphones with a spacing of $4.2\,\mathrm{cm}$. The microphone signals are computed by convolving \acp{RIR} measured in a living room environment with male and female speech signals and adding white Gaussian noise such that an \ac{SNR} of $30\,\mathrm{dB}$ at the microphones is obtained. \Walter{Two enclosures are considered in the following: Room 1 with a reverberation time of $T_{60} = 0.2\,\mathrm{s}$ and Room 2 with $T_{60} = 0.4\,\mathrm{s}$.} We placed $Q = 8$ acoustic sources at $1\,\mathrm{m}$, $2\,\mathrm{m}$ and $4\,\mathrm{m}$ distance and at different angles relative to the array for measuring the \acp{RIR} (see Fig.~\ref{fig:setup} for an illustration of the geometric setup of the measurements). All sources and microphones have been placed at the same height of $1.4\,\mathrm{m}$. The microphone signals are computed from a set of $4$ female and $4$ male speech signals of $20\,\mathrm{s}$ duration at a sampling frequency of $16\,\mathrm{kHz}$. The microphone signals are transformed into the \ac{STFT} domain using a von Hann window of \Walter{length} $2048$ and $50\%$ overlap. For the \ac{SOI} source models, we set $\beta = 1$ \Walter{in \eqref{eq:weightingFactor_IVA} and \eqref{eq:weightingFactor_NMF}}. \WalterThree{The performance of the investigated methods is measured in terms of the improvement (denoted by $\Delta$) of the \ac{SDR}, \ac{SIR} \WalterFour{and} \ac{SAR} \WalterFour{\cite{vincent_performance_2006}} w.r.t. the unprocessed microphone signals\WalterFour{, respectively,} and \WalterFour{in terms of} averaged runtime per iteration for all 20 permutations of the source signals.}

In the following, we aim at extracting a source $\SrcIdx$ (see Fig.~\ref{fig:setup}) out of the reverberant mixture of all sources. To obtain \Walter{representative} results, we repeat the experiment $20$ times \Walter{and permute} the \Walter{positions} of \Walter{the} speech \Walter{sources} in each \Walter{trial}. The performance of the algorithms is assessed by using the \Walter{improvement \WalterTwo{for} the} measures proposed by \cite{vincent_performance_2006}, where the separation of the \ac{SOI} from the mixture of all other signals is evaluated. The user-defined parameters are chosen for each algorithmic variant separately by a parameter sweep such that the best results are obtained on average for the extraction of source $\SrcIdx = 2$ for all $20$ permutations \Walter{(the choice of $\SrcIdx = 2$ is arbitrary here)}. Furthermore, the \Walter{parameters} have been chosen such that the outer permutation has been resolved, i.e., the desired source signal indeed appeared at the selected output channel. \Walter{The weighting parameters $\lambda$ and $\gamma$ have chosen to be equal for all frequency bins and channels.} The obtained parameters are summarized in Tab.~\ref{tab:parameters}.
\begin{figure}
	\flushright
	\setlength\fwidth{0.42\textwidth}
%
%
\begin{tikzpicture}

\begin{axis}[%
width=1\fwidth,
height=0.5\fwidth,
at={(0\fwidth,0\fwidth)},
scale only axis,
xmin=2,
xmax=13,
xlabel style={font=\color{white!15!black}},
xlabel={Algorithms $\rightarrow$},
ymin=-10,
ymax=8,
ylabel style={font=\color{white!15!black}},
ylabel={$\mathrm{dB}$ $\rightarrow$},
axis background/.style={fill=white},
xmajorgrids,
ymajorgrids,
legend style={at = {(1,0.15)},legend cell align=left, align=left, draw=white!15!black}, legend columns = 5
]

\node[fill = white] at (axis cs: 6,-6) {$\beta = $};
\node[fill = white] at (axis cs: 3.1,6.2) {\boxed{\textbf{$\Delta$SDR}}};

\addplot [color=black, mark=x, mark options={solid, black},line width = 1pt, mark size = 3]
table[row sep=crcr]{%
	2	1.99\\
	3	1.8192\\
	4	-10\\
	5	1.8597\\
	6	2.1451\\
	7	0.53456\\
};
\addlegendentry{$0$}

\addplot [color=black, dashdotted, mark=asterisk, mark options={solid, black},line width = 1pt, mark size = 3]
table[row sep=crcr]{%
	2	1.8487\\
	3	1.8976\\
	4	4.3385\\
	5	1.8606\\
	6	2.2467\\
	7	0.27377\\
	8	7.1531\\
	9	4.5522\\
	10	-0.8684\\
	11	5.5075\\
	12	1.9315\\
	13	4.8894\\
};
\addlegendentry{$0.5$}

\addplot [color=black, dashed, mark=o, mark options={solid, black},line width = 1pt, mark size = 3]
table[row sep=crcr]{%
	2	3.4864\\
	3	2.3316\\
	4	5.5198\\
	5	2.576\\
	6	2.3704\\
	7	4.24439\\
	8	7.0738\\
	9	6.9803\\
	10	4.4181\\
	11	3.6704\\
	12	2.4482\\
	13	5.4146\\
};
\addlegendentry{$1$}

\addplot [color=black, dotted, mark=square, mark options={solid, black},line width = 1pt, mark size = 3]
table[row sep=crcr]{%
	2	6.3597\\
	3	3.9432\\
	4	-1.6201\\
	5	5.6467\\
	6	3.4269\\
	7	1.4193\\
	8	4.2162\\
	9	3.6593\\
	10	0.0143\\
	11	2.144\\
	12	-0.20679\\
	13	1.5371\\
};
\addlegendentry{$1.5$}

\addplot [color=black, dotted, mark=diamond, mark options={solid, black},line width = 1pt, mark size = 3]
table[row sep=crcr]{%
	2	1.9419\\
	3	-2.2968\\
	4	-6.514\\
	5	3.9704\\
	6	-2.167\\
	7	-0.66383\\
	8	-0.72705\\
	9	-1.4134\\
	10	-2.3441\\
	11	0.96598\\
	12	-1.0897\\
	13	-0.23359\\
};
\addlegendentry{$2$}

\end{axis}
\end{tikzpicture}%
	\caption{Influence of the shape parameter $\beta$ of the \ac{SOI} model on the performance of \mbox{Methods 2-13} in terms of \ac{SDR} improvement.}
	\label{fig:SDR_beta}
\end{figure}
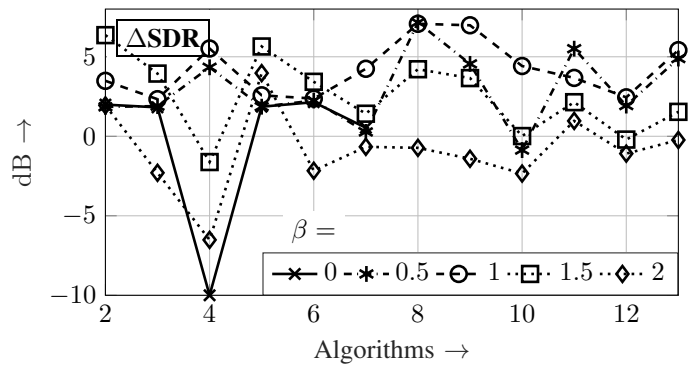
\subsection{Target Direction and Acoustic Environment}
The influence of different target \acp{DOA} (corresponding to sources $\SrcIdx = 1,2,3$) and of different acoustic environments is investigated in the following. To this end, the geometric setup, corresponding to Fig.~\ref{fig:setup}, is used in \Walter{the two} different rooms \Walter{described above} for measuring the \acp{RIR} and for each of these acoustic conditions source $\SrcIdx = 1,2,3$ is extracted. This experiment is again repeated for 20 permutations of the association between source positions and source signals and the median of the results is taken as a statistic, which is presented in Fig.~\ref{fig:differentScenarios}. The results of Room 1 are depicted as solid lines, the results of Room 2 as dashed lines. First of all, it can be seen that the extraction of source $\SrcIdx=3$ yielded the best results in terms of \ac{SDR} \Walter{improvement} for most algorithms, which may be explained by the geometric setup in which not many sources are contained in the angular region of source $\SrcIdx=3$. Furthermore, the performance of all algorithms degrades for Room 2, which has a higher reverberation time. This effect is typical for algorithms which perform spatial filtering. Also the assumption of free-field propagation used for the construction of the priors is violated for an increasing reverberation time. \Walter{While} the performance of most of the algorithms dropped only slightly, for the Methods 4, 7, 10, 13 a large drop can be observed. These methods have in common that they rely on the prior \eqref{eq:directional_prior} or \eqref{eq:directional_prior_BG} steering a spatial null. This spatial null constraint is imposed on all channels but one, instead of the priors steering a spatial one, which just impose a constraint on a single channel. As the free-field assumption is violated for increasing reverberation time, this has a larger effect on the methods using a prior steering a spatial null as this violated assumption is used multiple times. However, even for the methods with the large drop in the performance measures, \ac{SIR} improvement is achieved.

\subsection{Runtime, Source Models and SNR}
In terms of average runtime per iteration, Method 1 and 5-7 cause the lowest computational costs, followed by Methods 11-13. Hereby, the computational efficiency of the Methods 5-7 and 11-13 \Walter{results} from the usage of a \ac{BG} model. The computational cost of the Methods 2-4 and 8-10 is much higher than their counterparts using a \ac{BG} model. In terms of computational efforts to be spent until convergence, the gradient-based Method 1 is computationally much more costly as the number of iterations until convergence is much larger (about the factor $20-25$) than for the \ac{IP}-based methods.

\begin{figure}
	\flushright
	\setlength\fwidth{0.42\textwidth}
%
%
\begin{tikzpicture}

\begin{axis}[%
width=1\fwidth,
height=0.5\fwidth,
at={(0\fwidth,0\fwidth)},
scale only axis,
xmin=1,
xmax=13,
xlabel style={font=\color{white!15!black}},
xlabel={Algorithms $\rightarrow$},
ylabel style={font=\color{white!15!black}},
ylabel={$\mathrm{dB}\,\rightarrow$},
ymin=-3,
ymax=7.5,
axis background/.style={fill=white},
xmajorgrids,
ymajorgrids,
legend style={at={(0.457,0.275)},legend cell align=left, align=left, draw=white!15!black}, legend columns = 2
]

\node[fill = white] at (axis cs: 2.2,0.4) {$\mathrm{SNR} = $};
\node[fill = white] at (axis cs: 2.3,6.3) {\boxed{\textbf{$\Delta$SDR}}};

\addplot [color=black, mark=x, mark options={solid, black},line width = 1pt, mark size = 3]
table[row sep=crcr]{%
	1	2.2409\\
	2	3.3462\\
	3	2.1646\\
	4	3.7049\\
	5	2.0706\\
	6	2.2621\\
	7	3.4117\\
	8	3.942\\
	9	-0.55067\\
	10	-2.722\\
11	2.5181\\
12	2.5664\\
13	2.6773\\
};
\addlegendentry{$0\,\mathrm{dB}$}

\addplot [color=black, dashdotted, mark=asterisk, mark options={solid, black},line width = 1pt, mark size = 3]
table[row sep=crcr]{%
	1   2.8276\\
	2	3.3995\\
	3	2.4586\\
	4	5.0782\\
	5	2.6905\\
	6	2.5707\\
	7	5.145\\
	8	5.7006\\
	9	4.0484\\
	10	-0.5044\\
11	2.2008\\
12	2.7463\\
13	4.9888\\
};
\addlegendentry{$10\,\mathrm{dB}$}

\addplot [color=black, dashed, mark=o, mark options={solid, black},line width = 1pt, mark size = 3]
table[row sep=crcr]{%
	1	3.6932\\
	2	3.375\\
	3	2.3158\\
	4	5.4094\\
	5	2.4829\\
	6	2.2812\\
	7	5.6065\\
	8	6.731\\
	9	6.4714\\
	10	1.3713\\
11	4.3761\\
12	2.9546\\
13	6.2223\\
};
\addlegendentry{$20\,\mathrm{dB}$}

\addplot [color=black, dotted, mark=square, mark options={solid, black},line width = 1pt, mark size = 3]
table[row sep=crcr]{%
	1	3.928\\
	2	3.4864\\
	3	2.3351\\
	4	4.9158\\
	5	2.4733\\
	6	2.3775\\
	7	5.3511\\
	8	7.0738\\
	9	6.9803\\
	10	3.4181\\
11	3.9165\\
12	3.1935\\
13	5.9936\\
};
\addlegendentry{$30\,\mathrm{dB}$}
\end{axis}
\end{tikzpicture}%
	\caption{Influence of different noise levels on the discussed algorithmic variants in terms of \ac{SDR} improvement.}
	\label{fig:differentSNRs}
\end{figure}
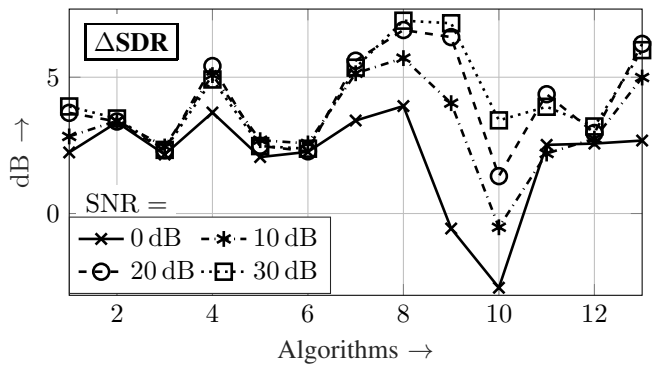
The influence of the number of bases $\BasisIdxMax$ for the Methods 8-13 relying on an \ac{NMF} source model is shown in Fig.~\ref{fig:num_bases}. It can be seen that for all methods $\BasisIdxMax = 2$ basis vectors provide satisfying results \WalterTwo{(see also, e.g., \cite{kitamura_determined_2016})}.

The influence of the shape factor $\beta$ of the \ac{SOI} models is discussed in terms of achieved \ac{SDR} improvement in Fig.~\ref{fig:SDR_beta}. The values $\beta = 0,0.5,1,1.5,2$ have been evaluated here (for the \ac{NMF}-based methods $\beta = 0$ is not evaluated as this would correspond to $\phi(r_{\ChannelIdx,\BlockIdx,\FreqIdx}) = 1$ $\forall \BlockIdx,\FreqIdx,\ChannelIdx$), where the value $\beta = 1$ corresponds to a Laplacian distribution and $\beta = 2$ to the time-varying Gaussian distribution \eqref{eq:SOI_model_timevaryingGaussian} w.r.t. the \ac{IVA} \ac{SOI} models. In case of the \ac{NMF} \ac{SOI} model, a time-varying Gaussian \ac{SOI} model is obtained for $\beta = 1$. \Walter{Inspection} of Fig.~\ref{fig:SDR_beta} shows that a choice of $\beta = 1$ yields good results for all algorithms. For some algorithmic variants the values of $\beta = 0.5$ or $\beta = 1.5$ are slightly better. In all cases, we obtain for the choice of $\beta = 0$ or $\beta = 2$ worse results. This is especially severe for Method 4, which relies on a prior steering a spatial one based on \eqref{eq:directional_prior}. 

The performance of the discussed algorithmic variants w.r.t. varying noise levels is shown in Fig.~\ref{fig:differentSNRs}. Here, we varied the additive noise, such that an \ac{SNR} of $0\,\mathrm{dB}$, $10\,\mathrm{dB}$, $20\,\mathrm{dB}$ and $30\,\mathrm{dB}$ is achieved at the microphones. \Walter{Unsurprisingly,} for an \ac{SNR} of $0\,\mathrm{dB}$ all algorithms produce the worst results. For the other noise levels, a detrimental effect due to the additive noise can be observed for the algorithms relying on an \ac{NMF} \ac{SOI} model, whereas the other methods are only slightly affected by the noise level. The detrimental effect of the increasing noise level is especially severe for Methods 8, 9, 10, which are using an \ac{NMF} source model and no \ac{BG} model.


\subsection{\Walter{Summary}}
\Walter{In this experimental study, we discussed different algorithms based on \ac{IVA} for source extraction, where the desired source is selected by a spatial constraint. In general, Methods 8-13 based on an \ac{NMF} source model yielded better results than Methods 1-7 \WalterTwo{(see Fig.~\ref{fig:differentScenarios})}. As another general outcome, it can be observed that methods using a spatial null constraint \WalterTwo{degraded severely} for increasing reverberation time. The influence of varying noise levels was not severe for most \acp{SNR} \WalterTwo{(see Fig.~\ref{fig:differentSNRs})}. The methods based on \ac{IP} showed much lower computational complexity than the baseline using gradient descent \cite{vincent_geometrically_2015} \WalterTwo{(see Fig.~\ref{fig:differentScenarios})}. The computational complexity can \WalterTwo{be further reduced significantly} by the use of an \ac{BG} model without sacrificing performance.} \WalterTwo{By comparing the results shown in Fig.~\ref{fig:differentScenarios}, it can be seen \WalterThree{there is no single best-performing} algorithm: For the TVG/SG source model, the proposed Algorithms 4 and 7 relying on a prior steering a spatial \WalterThree{null} perform especially well for $T_{60} = 0.2\,\mathrm{s}$ and degrades for larger $T_{60}$. For the algorithmic variants relying on \WalterThree{an} \ac{NMF} source model, the baseline Method 8 and the proposed Method 9, both steering a spatial one, yield similar results in all cases. However, the average runtime per iteration is slightly lower for the proposed Method 9. The proposed \ac{BG}-based Methods 11-13 obtained for some acoustic setup very good results but degraded for $T_{60} = 0.4\,\mathrm{s}$.}
\section{Conclusion}
\label{sec:conslusion}
In this contribution, we presented a unifying and flexible \WalterTwo{generic} framework for \WalterTwo{systematic} incorporation of prior knowledge on the demixing filters for \ac{IVA}-based source separation algorithms. \Walter{\WalterTwo{The potential of the framework was demonstrated for} several exemplary priors \WalterTwo{representing} geometric prior knowledge.} \Walter{As another generalization,} a \ac{BG} model is incorporated into the framework, which allows for fast convergence of the corresponding algorithms at a low computational cost if the number of \acp{SOI} is smaller than the number of microphones, i.e., $\MicIdxMax>\ChannelIdxMax$. \WalterTwo{The derivation of update rules for the \ac{BG} filters from this perspective \WalterThree{had} not been considered so far in the literature.} For all proposed algorithmic variants, we derived stable and fast update rules \WalterTwo{\WalterThree{with} a low computational complexity} based on the \ac{MM} principle and the \ac{IP} \WalterThree{approach, even including most recently proposed update rules into the systematic framework.}

The efficacy of the proposed algorithmic variants \WalterTwo{for real-world applications} is demonstrated by experiments using measured \acp{RIR} and \WalterThree{by} \Walter{comparison with \WalterTwo{established state-of-the-art baseline algorithms.}}

\bibliographystyle{IEEEtran}
\bibliography{literature}

%
%
%
%
%
%
%

\end{document}